\documentclass[manuscript]{aastex}

\slugcomment{}

\shorttitle{new HST/ACS of NGC~4449}
\shortauthors{Annibali et al.}

\begin{document}


\title{The cluster population of the irregular galaxy NGC~4449 as seen by the
Hubble Advanced Camera for Surveys\footnote{Based on observations 
with the NASA/ESA {\it Hubble Space Telescope}    obtained at the 
Space Telescope Science Institute    which is operated  by AURA    Inc.   
for NASA under contract NAS5-26555.}}



\author{F. Annibali\altaffilmark{2}, M. Tosi\altaffilmark{2}, A. Aloisi\altaffilmark{3}, R.~P. van der Marel\altaffilmark{3}   
} 

\altaffiltext{2}{INAF-Osservatorio Astronomico di Bologna, Via Ranzani 1, I-40127 Bologna, Italy}

\altaffiltext{3}{Space Telescope Science Institute, 3700 San Martin Drive, Baltimore MD 21218, USA}



\begin{abstract}

We present a study of the star cluster population in the starburst irregular galaxy NGC~4449 based on B, V, I, and H$\alpha$ images  
taken with the Advanced Camera for Surveys on the Hubble Space Telescope. We derive the cluster properties such as size, ellipticity, and 
total magnitudes. 
Cluster ages and masses are derived fitting the observed spectral energy distributions with different population synthesis models. Our analysis is strongly affected by the age-metallicity 
degeneracy; however, if we assume a metallicity of $\sim$1/4 solar, as derived from spectroscopy of HII regions, we find that the clusters have ages distributed quite continuously over a Hubble time, and they have masses from $\sim10^3  M_{\odot}$ up to $\sim2\times10^6 M_{\odot}$, assuming a Salpeters' IMF down to 0.1 $M_{\odot}$. 
Young clusters are preferentially located in regions of young star formation, 
while old clusters are distributed over the whole NGC 4449 field of view, like the old stars (although we notice that 
some old clusters follow linear structures, possibly a reflection of past satellite accretion).
The high SF activity in NGC~4449 is confirmed by its specific frequency of young massive clusters, higher than the average value found in nearby spirals and in the LMC (but lower than in other starburst dwarfs such as 
NGC~1705 and NGC~1569), and by the flat slope of the cluster luminosity function ($dN(L_V)\propto L_V^{-1.5} dL$ for clusters younger than 1 Gyr). We use the upper envelope of the cluster log(mass) versus log(age) distribution to quantify cluster disruption, and do not find evidence 
for the high (90\%) long-term infant mortality found by some studies. For the red clusters, we find correlations between size, ellipticity, 
luminosity and mass: brighter and more massive clusters tend to be more compact, and brighter clusters tend to be also more elliptical.

\end{abstract}


\keywords{galaxies: dwarf --- galaxies: individual (NGC~4449) 
---galaxies: irregular ---galaxies: star clusters: general 
---galaxies: starburst }



\section{Introduction}

Star clusters are present in all types of galaxies, from the earliest to the
latest morphological types, from the quietest to the most turbulent
ones, such as merging and starburst systems. They are suggested to be the birth 
site of many (possibly most) stars, and provide an excellent tool to study 
the parent galaxy evolution through a long range of epochs. For these reasons,
the interest in star cluster studies never fades. While Galactic star clusters
were already studied a century ago, extragalactic ones have required  
 the superb spatial resolution of the Hubble Space Telescope (HST) to receive
the deserved attention. Nowadays people use HST to (at least partially) 
resolve both old globulars (GCs) and young and intermediate-age clusters up to 
distances of a few Mpc. Thanks to HST, special interest has been focussed in the 
last two decades to the young massive clusters typical of star forming galaxies,
including the so-called Super Star Clusters - SSCs - (e.g.
\citet{oco94,bill02,lar04,whi05,sab07,ann09,lar09,mora09}). 
These objects have masses typical of globular clusters and may thus be 
the young counterpart of the classical fossil records of the earliest localized 
star formation activity. If globular clusters could still be formed today in some 
environments, their study would provide a direct insight into the conditions that were 
present in the early days of galaxy formation, when the globular clusters that we see 
today in the halos were formed.

In this paper we present new data on the star clusters
in the starburst irregular galaxy NGC~4449. 
NGC~4449 is known to host a young ($\sim$ 6-15 Myr) central super star cluster \citep{bok01,gel01}, 
numerous star clusters  ($\sim 60$, \cite{gel01}) with ages up to 1 Gyr and possibly older, and at least 13  embedded massive 
clusters detected in the radio with ages $\lesssim$5 Myr \citep{reines08}.
Our analysis is based on our HST data acquired with the Wide Field Channel (WFC) of the Advanced Camera for Surveys (ACS).

The Magellanic irregular galaxy NGC~4449
($\alpha_{2000} =Ê12^h 28^m 11^{s}.9$    
$\delta_{2000} =Ê+  44^{\circ} 05^{'} 40^{"}$    $l=136.84$ 
and $b=72.4$)    
at a distance of $3.82 \pm 0.18$ Mpc
(\cite{ann08}, hereafter A08) is one of the best studied and spectacular
nearby starbursts. 
Its size and metallicity 
are similar to those of the Large Magellanic Cloud (LMC), 
while its integrated magnitude $M_B
= -18.2$ (\cite{hunter99}, A08) is $\approx$ 1.4 times brighter than the LMC's. 
Also its SFR, averaged over a Hubble time, is $\sim 0.2$ M$_{\odot}$ yr$^{-1}$,  twice
that of the LMC, while its current SFR is as high as $\sim 1.5$ M$_{\odot}$ yr$^{-1}$ \citep{th87}. 
It is indeed one of the most luminous and active irregular galaxies, as confirmed also by the recent study by
\cite{mcq10} on the star formation histories (SFHs) of 18 nearby
starburst galaxies, of which NGC~4449 turns out to be the strongest one.

Data acquisition and reduction are summarized in Section~\ref{data_reduction}, while the cluster
analysis and the artificial cluster experiments are described in Section~\ref{cluster_ana}. The results (luminosity functions, ages and masses, cluster classification, correlations between cluster properties) are presented in Section~\ref{resu}. 
In Section~\ref{cluster_disruption} we investigate cluster disruption.
The final results and discussion are presented in Section~\ref{discussion}.

\section{Observations and data reduction\label{data_reduction}}

Here we briefly recall the adopted observing strategy 
and main steps involved in the data reduction process, 
which were fully described in A08.

The observations were performed in November 2005 with
the ACS/WFC using the F435W (B), F555W (V), and F814W (I) 
broad-band filters, and the F658N (H$\alpha$)
narrow-band filter (GO program 10585, PI Aloisi).
We had two different pointings
in a rectangular shape along the major axis of the galaxy.
Each pointing was organized with a 4 - exposure 
half $+$ integer pixel dither pattern
with the following offsets in arcseconds     
(0, 0) for exposure 1, (0.12, 0.08) for exposure 2,   
(0.25, 2.98) for exposure 3, and (0.37, 3.07)
for exposure 4. This dither pattern is suitable to remove
cosmic rays and hot/bad pixels, fill the gap between the
two CCDs of the ACS/WFC, and improve the PSF sampling.
Four exposures of $\sim$ 900 s, 600 s, 500 s and
90 s for each of the two fields were acquired for each of the B, V, I and 
H$\alpha$ filters, respectively.

For each filter, the eight frames (4 dithered exposures for each of the 2 fields), calibrated
through the most up-to-date version of the ACS calibration 
pipeline (CALACS), were co-added into a single mosaicked image using the
software package MULTIDRIZZLE \citep{Koe02}.
The MULTIDRIZZLE procedure also corrects
the ACS images for geometric distortion and
provides removal of cosmic rays and bad pixels.
The total field of view of the resampled 
mosaicked image is $\sim$ (380 $\times$ 200) arcsec$^2$    
with a pixel size of 0.035 \arcsec (0.7 times the 
original ACS/WFC pixel size). 

The total integration times are $\sim$ 3600 s, 2400 s,     
2000 s and 360 s for the B, V, I and H$\alpha$ images, respectively.
Only in a small region of overlap between the two pointings
($\sim$ (30 $\times$ 200) arcsec$^2$) the integration times are
twice as those listed above.

The source detection was performed with the DAOFIND task 
within IRAF\footnote{IRAF is distributed
by the National Optical Astronomy Observatories, which are operated by
AURA, Inc., under cooperative agreement with the National Science 
Foundation} above 3 times the background standard deviation, 
independently in the three bands. 
PSF-fitting photometry was performed with the
DAOPHOT package \citep{daophot} in the IRAF  environment.
The PSF was computed from the 
most isolated stars in the frame, and 
was modeled with an analytic Moffat function 
plus additive corrections derived from
the residuals of the fit to the PSF stars.
Then we ran  ALLSTAR to iteratively fit the PSF model to groups of stars, 
providing coordinates and magnitudes of the
individual stars, along with magnitude errors, 
$\chi^2$,  and sharpness.
In particular,  the {\it sharpness} parameter provides a measure
of the intrinsic size of the object with respect 
to the PSF. 
Sources with profile widths similar 
to the PSF's have sharpness of about zero, cosmic rays have sharpness less than zero, 
and extended sources have sharpness larger than zero.
The PSF FWHM is $\lesssim$ 0.1'', corresponding to $\sim$ 2 pc at NGC~4449' s distance.

The instrumental magnitudes were transformed into the 
HST VEGAMAG system following \citet{sir05}. 
Aperture corrections were calculated from isolated stars selected in our images.
Corrections for imperfect charge transfer efficiency (CTE), 
applied following the formulation of \cite{cte}, are negligible for the brightest stars,
but can be as high as $\sim$ 0.1 mag for the faintest stars.

The B,V and I catalogues were cross-correlated
with the requirement of a spatial offset
smaller than 1 pixel between the positions of the stars
in the different frames.
This led to 299,115 objects having a measured magnitude 
in both B and V, 402,136 objects in V and I, and 213,187
objects photometred in all the three bands simultaneously.

\section{Cluster analysis \label{cluster_ana}}

\subsection{Selection \label{selection}}

\begin{figure}
\epsscale{1.}
\plotone{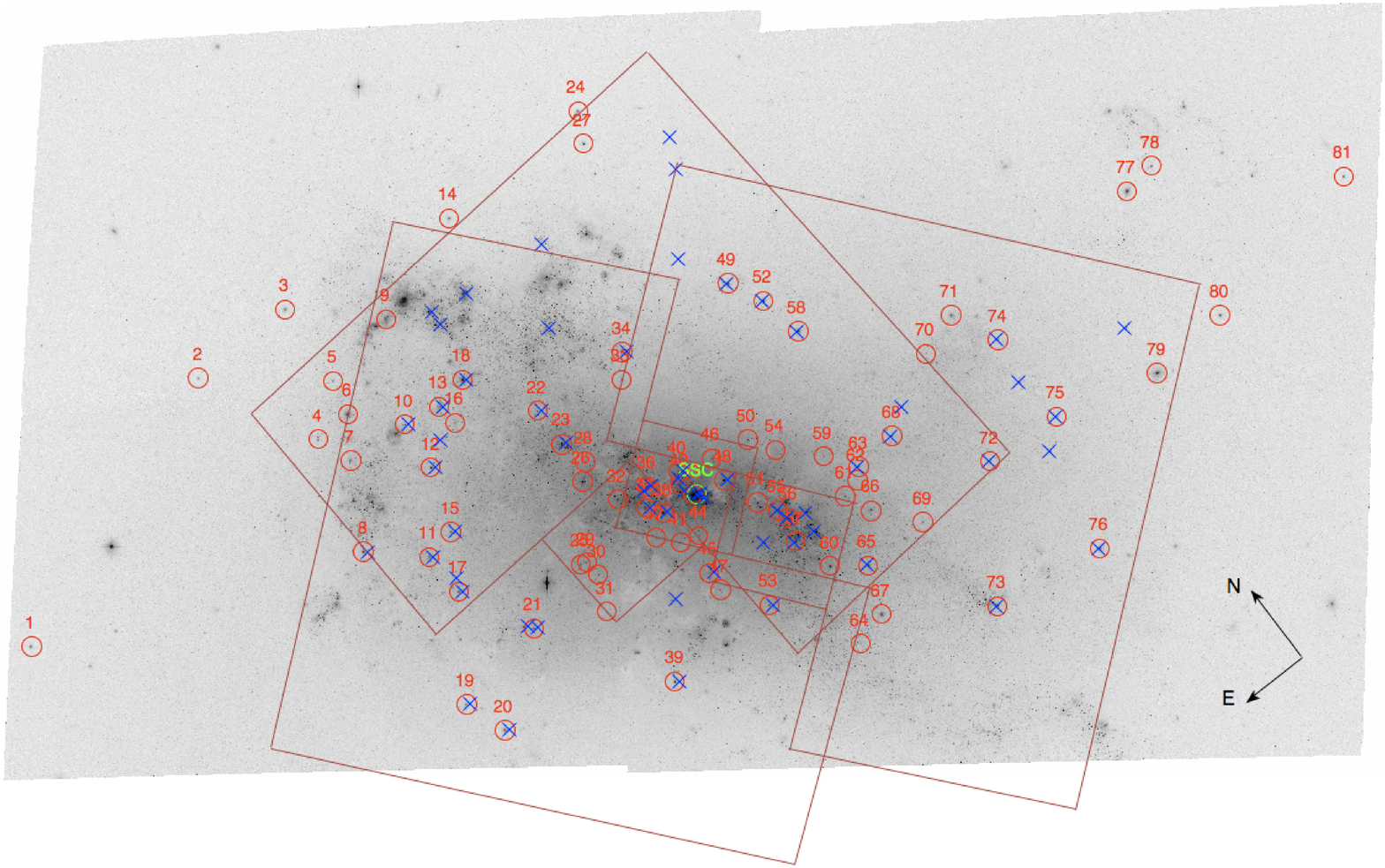}
\caption{Mosaicked ACS F555W image (380 $\times$ 200 arcsec$^2$) with superimposed the 81 candidate star clusters (red circles). 
The footprint of the WFPC2 images used by \cite{gel01} and their candidate clusters (blue crosses) are shown for comparison. \label{image_clusters}}
\label{image}
\end{figure}

Cluster candidates were selected from the DAOPHOT photometric catalog adopting a 
magnitude cut  of ${\rm m_{F814W} }\la 22$ \footnote{At fainter magnitudes, contamination by NGC 4449~stars, background galaxies, and blends becomes severe. Also, our completeness for cluster detection drops significantly, as quantified in 
Section~\ref{artificial}} and requiring {\it sharpness$_{DAO}$} $>$0.5 in all three F435W, F555W, and F814W filters.
After this preliminary selection, we visually inspected the objects in all the three images looking for centrally concentrated clusters.
At a visual inspection, some of the objects turned out to be blends of two or more point-like sources,    
while others appeared as bona fide  background galaxies, with irregular fuzzy structures, spiral arms, or disky features.    
These objects were removed from the cluster candidate lists. 
We ended up with a final list of 81 candidates, 
whose coordinates are provided in Cols.~2 and 3 of Table~1.
Their location in the NGC~4449 field of view is shown in Fig.~\ref{image_clusters}.
Background elliptical galaxies may still be present in the candidate list, but 
the fact that the candidates do not appear uniformly distributed over our field of view suggests  
that the majority of them are indeed clusters belonging to NGC~4449: 
in fact, most of them are either concentrated in the galaxy 
central bar, or close to the S-shaped filaments delineated by stars 
younger that 10 Myr and by the H$_{\alpha}$ emission
(Figs. 19 and 20 in A08).

The clusters are individually shown in Fig.~\ref{clustera}.
We notice that a significant number of them are partially resolved into stars in our images, 
which further confirms that the contamination from background sources is minimal.
We also show in Fig.~\ref{clu_ha} color-composite images of the clusters associated with H$\alpha$ emission, exhibiting a variety of 
ionized gas morphologies.  Our F658N images are quite shallow, thus it is possible that we are missing clusters with very faint ionized gas emission.

\cite{gel01} performed a study of the star cluster population in NGC~4449 using WFPC2 images.
Their data cover a field of view of $\sim$ (240$\times$ 200) arcsec$^2$, smaller than ours   
$\sim$ (380$\times$ 200) arcsec$^2$ (see Fig.~\ref{image_clusters}).  
Besides the SSC at the center of NGC~4449, they identified 60 candidate compact star clusters,  37 of which are 
in common with us. We visually inspected in our images the objects classified as clusters by  \cite{gel01} but not by us.
Some of them appear extended in F435W or F555W, but are resolved into two or more stars in F814W. Others appear more 
similar to stellar associations (but we can not distinguish between ``open clusters'' or chance superpositions of unaffiliated stars) rather than to compact clusters. We show in Fig.~\ref{2exgel} two examples of objects classified by \cite{gel01} as clusters, but considered by us to be ``stellar associations''.
For the 37 clusters in common, we provide in Table~1 the ID assigned by \cite{gel01}. Our ACS images are saturated in the center of the galaxy, preventing a study of the SSC in NGC~4449's center.

\begin{deluxetable}{llllllllllll}
\tabletypesize{\scriptsize}
\rotate
\tablewidth{0pt}
\tablecolumns{12}
\tablecaption{Cluster properties}
\tablehead{
\colhead{ID\tablenotemark{a}} & \colhead{R.A.} & \colhead{Dec} & \colhead{R$_e$\tablenotemark{b}} & \colhead{$\epsilon$\tablenotemark{c}}  & \colhead{$m_{F435W}$\tablenotemark{d}} & \colhead{$m_{F555W}$\tablenotemark{d}}    & \colhead{$m_{F814W}$\tablenotemark{d}} & \colhead{$M_V$\tablenotemark{e}} & \colhead{Age\tablenotemark{f}} & \colhead{Mass\tablenotemark{f}} & \colhead{$M_V$ (10 Myr)\tablenotemark{g}} \\
\colhead{} & \colhead{J2000} & \colhead{J2000} & \colhead{''(pc)}  & \colhead{} & \colhead{} & \colhead{}    & \colhead{} & \colhead{} & \colhead{Myr} & \colhead{$10^5 M_{\odot}$} & \colhead{}   
}
\startdata
SSC &    12 28   11.13  & 44  05  37.42   & $-$  & $-$  & $-$  & $-$  & $-$ & $-$ & $-$  & $-$ & $-$ \\
    1  &    12  28  27.04  & 44  07  00.54   &  0.17     (3.2) &  0.07   &  21.24$\pm$0.01   &  20.49$\pm$0.01  &  19.41$\pm$0.01  &   -7.56$\pm$0.15  &  $1441_{-4}^{+4}$          &  $0.49_{-0.01}^{+0.01}$  & $-11.03_{-0.15}^{+0.15}$   \\      
     2  &    12  28  19.31  & 44  07  29.72   &  0.21     (3.8) &  0.08   &  22.24$\pm$0.03   &  21.52$\pm$0.02  &  20.51$\pm$0.03  &   -6.53$\pm$0.15  &  $3601_{-673}^{+285}$      &  $0.58_{-0.08}^{+0.03}$  & $-11.21_{-0.17}^{+0.24}$   \\  
     3  &    12  28  16.44  & 44  07  29.49   &  0.49     (9.0) &  0.04   &  22.42$\pm$0.08   &  21.63$\pm$0.07  &  20.56$\pm$0.09  &   -6.42$\pm$0.17  &  $7035_{-349}^{+326}$      &  $0.90_{-0.03}^{+0.03}$  & $-11.73_{-0.17}^{+0.17}$   \\      
     4  &    12  28  17.88  & 44  06  55.41   &  0.21     (3.9) &  0.05   &  20.90$\pm$0.01   &  20.52$\pm$0.01  &  19.81$\pm$0.02  &   -7.50$\pm$0.15  &  $609_{-45}^{+58}$         &  $0.37_{-0.02}^{+0.02}$  & $-10.66_{-0.16}^{+0.16}$   \\     
     5  &    12  28  16.62  & 44  07  05.62   &  0.12     (2.1) &  0.11   &  21.26$\pm$0.01   &  20.97$\pm$0.01  &  20.22$\pm$0.02  &   -7.06$\pm$0.15  &  $444_{-19}^{+24}$         &  $0.21_{-0.01}^{+0.01}$  & $ -9.98_{-0.16}^{+0.15}$   \\      
      6  &    12  28  16.87  & 44  06  55.72   &  0.33     (6.1) &  0.03   &  18.87$\pm$0.01   &  18.59$\pm$0.01  &  18.01$\pm$0.02  &   -9.42$\pm$0.15  &  $295_{-10}^{+10}$         &  $1.36_{-0.01}^{+0.01}$  & $-12.03_{-0.15}^{+0.15}$   \\      
      7  &    12  28  17.57  & 44  06  45.33   &  0.31     (5.8) &  0.09   &  20.95$\pm$0.04   &  20.19$\pm$0.04  &  19.00$\pm$0.03  &   -7.87$\pm$0.15  &  $1510_{-54}^{+2}$         &  $0.76_{-0.09}^{+0.01}$  & $-11.49_{-0.15}^{+0.24}$   \\      
      8$^{\ast}$(59)  &    12  28  18.79  & 44  06  23.08   &  0.20     (3.7) &  0.11   &  20.22$\pm$0.02   &  19.48$\pm$0.01  &  18.43$\pm$0.02 &   -8.57$\pm$0.15  &  $5181_{-660}^{+881}$     &  $5.12_{-0.42}^{+0.75}$  & $-13.59_{-0.22}^{+0.19}$   \\      
     9$^{\star}$  &    12  28  14.56  & 44  07  09.72   &  0.04     (0.7) &  0.15   &  19.02$\pm$0.01   &  18.92$\pm$0.01  &  18.64$\pm$0.01  &   -9.08$\pm$0.15  &  $ 83_{-28}^{+26}$         &  $0.50_{-0.10}^{+0.07}$  & $-10.87_{-0.22}^{+0.29}$   \\     
   10$^{\dagger \ast}$(51)$^{\star}$  & 12 28 15.89    & 44 06 43.54 &  0.07 (1.4) &  0.15   &  21.54$\pm$0.01   &  20.22$\pm$0.01  &  20.84$\pm$0.02  &   -7.70$\pm$0.15  &  --        &   --  & --   \\    
    11$^{\ast}$(58)  &    12  28  17.54  & 44  06  10.45   &  0.32     (6.0) &  0.08   &  20.41$\pm$0.03   &  20.21$\pm$0.04  &  19.69$\pm$0.07 &   -7.80$\pm$0.15  &  $ 259_{-1}^{+1}$          &  $0.30_{-0.01}^{+0.01}$  & $-10.35_{-0.15}^{+0.15}$   \\     
    12$^{\dagger \ast}$(52)$^{\star}$  & 12 28 16.05    & 44 06 29.64 &  0.08 (1.5) &  0.13   &  20.22$\pm$0.01   &  18.55$\pm$0.01  &  19.25$\pm$0.01  &   -9.37$\pm$0.15  &  --        &   --  & -- \\    
   13$^{\ast}$(49)$^{\star}$  &    12  28  14.90  & 44  06  41.39   &  0.09     (1.7) &  0.36   &  20.51$\pm$0.01   &  20.63$\pm$0.02  &  20.92$\pm$0.07 &   -7.33$\pm$0.15  &  $5_{-5}^{+1}$             &  $0.02_{-0.02}^{+0.01}$  & $ -7.06_{-0.18}^{+1.83}$   \\     
   14  &    12  28  11.65  & 44  07  20.74   &  0.08     (1.5) &  0.05   &  20.59$\pm$0.01   &  19.87$\pm$0.01  &  18.87$\pm$0.01  &   -8.18$\pm$0.15  &  $3240_{-326}^{+396}$      &  $2.44_{-0.19}^{+0.19}$    & $-12.77_{-0.18}^{+0.18}$   \\     
   15$^{\ast}$(54)  &    12  28  16.69  & 44  06  12.18   &  0.20     (3.7) &  0.04   &  20.25$\pm$0.02   &  20.01$\pm$0.02  &  19.39$\pm$0.04 &   -8.01$\pm$0.15  &  $296_{-1}^{+1}$           &  $0.38_{-0.01}^{+0.01}$  & $-10.62_{-0.15}^{+0.15}$   \\     
   16  &    12  28  14.84  & 44  06  35.16   &  0.26     (4.8) &  0.09   &  21.15$\pm$0.04   &  20.85$\pm$0.05  &  20.15$\pm$0.09  &   -7.18$\pm$0.16  &  $389_{-22}^{+60}$         &  $0.20_{-0.01}^{+0.02}$  & $-10.00_{-0.19}^{+0.17}$   \\     
   17$^{\ast}$(57)  &    12  28  17.51  & 44  05  57.78   &  0.17     (3.2) &  0.10   &  19.89$\pm$0.01   &  19.70$\pm$0.01  &  19.27$\pm$0.03 &   -8.30$\pm$0.15  &  $251_{-16}^{+3}$          &  $0.45_{-0.03}^{+0.01}$  & $-10.83_{-0.15}^{+0.16}$   \\     
   18$^{\ast}$(47)  &    12  28  13.99  & 44  06  43.21   &  0.16     (2.9) &  0.27   &  18.15$\pm$0.01   &  17.83$\pm$0.01  &  17.19$\pm$0.01 &  -10.19$\pm$0.15  &  $330_{-23}^{+20}$         &  $2.89_{-0.01}^{+0.03}$  & $-12.88_{-0.16}^{+0.16}$   \\     
   19$^{\ast}$(61)  &    12  28  19.19  & 44  05  32.00   &  0.29     (5.3) &  0.05   &  21.58$\pm$0.02   &  20.73$\pm$0.02  &  19.50$\pm$0.02 &   -7.34$\pm$0.15  &  $13197_{-1524}^{}$        &   $3.51_{-0.24}^{}$      & $-13.15_{-0.15}^{+0.18}$   \\     
   20$^{\ast}$(60)  &    12  28  18.82  & 44  05  19.61   &  0.48     (8.8) &  0.03   &  20.53$\pm$0.01   &  19.75$\pm$0.01  &  18.62$\pm$0.01 &   -8.30$\pm$0.15  &  $1450_{-4}^{+24}$         &   $0.93_{-0.01}^{+0.05}$ & $-11.73_{-0.17}^{+0.15}$   \\     
   21$^{\ast}$(53)  &    12  28  16.58  & 44  05  36.64   &  0.19     (3.5) &  0.02   &  20.33$\pm$0.02   &  19.59$\pm$0.01  &  18.54$\pm$0.02 &   -8.46$\pm$0.15  &  $5010_{-649}^{+1001}$    &   $4.53_{-0.39}^{+0.74}$ & $-13.45_{-0.24}^{+0.19}$   \\     
   22$^{\ast}$(42)  &    12  28  12.99  & 44  06  23.42   &  0.19     (3.5) &  0.12   &  20.32$\pm$0.02   &  20.07$\pm$0.02  &  19.57$\pm$0.04 &   -7.94$\pm$0.15  &  $267_{-9}^{+10}$         &   $0.33_{-0.01}^{+0.01}$ & $-10.50_{-0.15}^{+0.15}$   \\    
   23$^{\ast}$(43)  &    12  28  13.03  & 44  06  12.01   &  0.12     (2.1) &  0.20   &  18.25$\pm$0.02   &  18.21$\pm$0.02  &  17.71$\pm$0.02 &   -9.80$\pm$0.15  &  $28_{-2}^{+2}$            &   $0.56_{-0.01}^{+0.01}$ & $-10.85_{-0.16}^{+0.16}$   \\     
   24  &    12  28  07.32  & 44  07  21.54   &  0.30     (5.5) &  0.09   &  21.23$\pm$0.01   &  20.42$\pm$0.01  &  19.34$\pm$0.02  &   -7.64$\pm$0.15  &  $7675_{-63}^{+64}$        &   $2.92_{-0.01}^{+0.01}$ & $-13.02_{-0.15}^{+0.15}$   \\     
   25  &    12  28  14.61  & 44  05  42.47   &  0.31     (5.7) &  0.18   &  20.49$\pm$0.03   &  20.22$\pm$0.04  &  19.60$\pm$0.06  &   -7.80$\pm$0.15  &  $303_{-3}^{+5}$           &   $0.31_{-0.01}^{+0.01}$ & $-10.42_{-0.15}^{+0.15}$   \\     
   26  &    12  28  13.23  & 44  06  00.10   &  0.21     (3.9) &  0.27   &  18.40$\pm$0.03   &  18.32$\pm$0.03  &  18.04$\pm$0.05  &   -9.68$\pm$0.15  &  $ 9_{-1}^{+1}$            &   $0.17_{-0.01}^{+0.01}$ & $ -9.56_{-0.16}^{+0.16}$   \\     
   27  &    12  28  07.72  & 44  07  13.42   &  0.10     (1.8) &  0.07   &  20.75$\pm$0.01   &  19.93$\pm$0.01  &  18.81$\pm$0.01  &   -8.12$\pm$0.15  &  $9264_{-747}^{+1021}$    &   $5.36_{-0.24}^{+0.33}$ & $-13.66_{-0.17}^{+0.16}$   \\     
   28  &    12  28  12.85  & 44  06  04.01   &  0.27     (5.1) &  0.13   &  20.46$\pm$0.03   &  20.16$\pm$0.03  &  19.49$\pm$0.04  &   -7.87$\pm$0.15  &  $ 332_{-11}^{+17}$        &   $0.35_{-0.01}^{+0.01}$ & $-10.56_{-0.16}^{+0.15}$   \\    
   29  &    12  28  14.45  & 44  05  41.87   &  0.12     (2.2) &  0.09   &  20.82$\pm$0.02   &  20.55$\pm$0.02  &  19.71$\pm$0.03  &   -7.49$\pm$0.15  &  $ 363_{-14}^{+276}$       &   $0.29_{-0.01}^{+0.12}$ & $-10.26_{-0.45}^{+0.15}$   \\    
   30  &    12  28  14.41  & 44  05  37.08   &  0.15     (2.7) &  0.08   &  21.05$\pm$0.02   &  20.74$\pm$0.02  &  20.13$\pm$0.06  &   -7.28$\pm$0.15  &  $ 317_{-22}^{+30}$        &   $0.19_{-0.01}^{+0.01}$ & $ -9.94_{-0.17}^{+0.16}$   \\    
   31  &    12  28  14.85  & 44  05  27.51   &  0.07     (1.3) &  0.10   &  21.70$\pm$0.02   &  21.41$\pm$0.02  &  20.76$\pm$0.04  &   -6.61$\pm$0.15  &  $ 321_{-12}^{+14}$        &   $0.11_{-0.01}^{+0.01}$ & $ -9.28_{-0.15}^{+0.15}$   \\    
   32  &    12  28  12.81  & 44  05  50.06   &  0.38     (7.1) &  0.10   &  20.66$\pm$0.27   &  20.31$\pm$0.20  &  19.68$\pm$0.20  &   -7.71$\pm$0.25  &  $ 327_{-26}^{+24}$       &   $0.29_{-0.01}^{+0.09}$ & $-10.40_{-0.25}^{+0.25}$   \\    
   33  &    12  28  10.78  & 44  06  15.48   &  0.35     (6.5) &  0.10   &  20.67$\pm$0.03   &  20.39$\pm$0.05  &  19.86$\pm$0.09  &   -7.63$\pm$0.16  &  $ 278_{-12}^{+19}$        &   $0.25_{-0.01}^{+0.01}$ & $-10.20_{-0.16}^{+0.16}$   \\    
   34$^{\ast}$(26)  &    12  28  10.32  & 44  06  21.37   &  0.13     (2.5) &  0.24   &  19.40$\pm$0.01   &  18.61$\pm$0.01  &  17.51$\pm$0.01 &   -9.45$\pm$0.15  &  $7758_{-489}^{+853}$     &   $15.92_{-0.61}^{+1.20}$& $-14.84_{-0.17}^{+0.16}$   \\    
   35$^{\ast}$(40)  &    12  28  12.37  & 44  05  43.21   &  0.12     (2.3) &  0.20   &  19.14$\pm$0.03   &  18.39$\pm$0.02  &  17.18$\pm$0.02 &   -9.68$\pm$0.15  &  $1510_{-23}^{+2}$         &   $4.04_{-0.23}^{+0.06}$ & $-13.30_{-0.15}^{+0.17}$   \\    
   36$^{\ast}$(36)  &    12  28  12.01  & 44  05  47.63   &  0.19     (3.5) &  0.34   &  18.63$\pm$0.03   &  18.43$\pm$0.03  &  17.61$\pm$0.03 &   -9.60$\pm$0.15  &  $ 363_{-19}^{+106}$       &   $2.02_{-0.04}^{+0.38}$ & $-12.37_{-0.25}^{+0.16}$   \\    
   37  &    12  28  12.65  & 44  05  35.14   &  0.30     (5.5) &  0.11   &  21.27$\pm$0.05   &  20.74$\pm$0.05  &  19.69$\pm$0.06  &   -7.31$\pm$0.16  &  $1550_{-7}^{+195}$       &   $0.60_{-0.04}^{+0.14}$ & $-11.18_{-0.24}^{+0.16}$   \\    
   38$^{\ast}$(37)  &    12  28  12.11  & 44  05  39.25   &  0.19     (3.5) &  0.23   &  18.01$\pm$0.02   &  17.68$\pm$0.01  &  16.63$\pm$0.02 &  -10.37$\pm$0.15  &  $1070_{-230}^{+228}$      &   $8.85_{-0.87}^{}$      & $-13.95_{-0.16}^{+0.24}$   \\    
   39$^{\ast}$(48)  &    12  28  14.60  & 44  05  00.43   &  0.09     (1.7) &  0.10   &  20.24$\pm$0.01   &  19.43$\pm$0.01  &  18.27$\pm$0.01 &   -8.63$\pm$0.15  &  $11033_{-1820}^{}$        &   $9.78_{-0.96}^{}$      & $-14.30_{-7.76}^{+0.20}$   \\    
   40$^{\ast}$(30)  &    12  28  11.12  & 44  05  45.56   &  0.09     (1.7) &  0.16   &  19.30$\pm$0.03   &  18.67$\pm$0.02  &  17.59$\pm$0.02 &   -9.38$\pm$0.15  &  $1537_{-2}^{+2}$          &   $3.64_{-0.09}^{+0.08}$ & $-13.18_{-0.15}^{+0.15}$   \\    
   41  &    12  28  12.25  & 44  05  29.89   &  0.24     (4.5) &  0.14   &  20.75$\pm$0.03   &  20.57$\pm$0.04  &  20.13$\pm$0.08  &   -7.44$\pm$0.15  &  $ 251_{-17}^{+3}$         &   $0.20_{-0.01}^{+0.01}$ & $ -9.97_{-0.15}^{+0.17}$   \\    
   42$^{\star}$  &    12  28  11.30  & 44  05  42.41   &  0.21     (3.8) &  0.10   &  20.02$\pm$0.10   &  19.77$\pm$0.10  &  19.09$\pm$0.12  &   -8.26$\pm$0.18  &  $ 320_{-2}^{+2}$          &   $0.50_{-0.01}^{+0.01}$ & $-10.92_{-0.18}^{+0.18}$   \\    
   43$^{\ast}$(31)$^{\star}$  &    12  28  11.22  & 44  05  38.06   &  0.11     (2.0) &  0.41   &  17.43$\pm$0.01   &  17.25$\pm$0.01  &  16.93$\pm$0.01 &  -10.75$\pm$0.15  &  $ 111_{-18}^{+5}$        &   $2.68_{-0.15}^{+0.01}$ & $-12.72_{-0.15}^{+0.18}$   \\    
   44 &    12  28  11.74  & 44  05  28.03   &  0.36     (6.7) &  0.10   &  20.50$\pm$0.26   &  20.16$\pm$0.18  &  19.44$\pm$0.14  &   -7.87$\pm$0.23  &  $ 537_{-55}^{+34}$        &   $0.48_{-0.03}^{+0.02}$ & $-10.94_{-0.24}^{+0.24}$   \\    
   45$^{\ast}$(38)  &    12  28  12.15  & 44  05  18.15   &  0.13     (2.4) &  0.17   &  20.61$\pm$0.02   &  20.43$\pm$0.02  &  20.05$\pm$0.05 &   -7.58$\pm$0.15  &  $ 41_{-5}^{+6}$          &   $0.08_{-0.01}^{+0.01}$ & $ -8.89_{-0.17}^{+0.16}$   \\    
   46  &    12  28  10.24  & 44  05  42.57   &  0.18     (3.4) &  0.06   &  20.65$\pm$0.15   &  19.92$\pm$0.10  &  18.79$\pm$0.08  &   -8.14$\pm$0.18  &  $1445_{-1}^{+1}$          &   $0.83_{-0.01}^{+0.01}$ & $-11.59_{-0.18}^{+0.18}$   \\   
   47  &    12  28  12.21  & 44  05  12.60   &  0.20     (3.6) &  0.17   &  21.22$\pm$0.04   &  20.91$\pm$0.04  &  20.30$\pm$0.08  &   -7.11$\pm$0.16  &  $ 311_{-18}^{+22}$        &   $0.16_{-0.01}^{+0.01}$ & $ -9.75_{-0.16}^{+0.16}$   \\    
   48$^{\ast}$(27)  &    12  28  10.35  & 44  05  35.79   &  0.19     (3.5) &  0.04   &  18.68$\pm$0.03   &  18.27$\pm$0.03  &  17.43$\pm$0.02 &   -9.77$\pm$0.15  &  $ 891_{-35}^{+40}$        &   $3.95_{-0.07}^{+0.09}$ & $-13.21_{-0.15}^{+0.15}$   \\    
   49$^{\dagger \ast}$(16)$^{\star}$  & 12 28 07.07    & 44 06 17.82 &  0.12 (2.2) &  0.18   &  20.47$\pm$0.01   &  20.16$\pm$0.01  &  20.48$\pm$0.02  &   -7.79$\pm$0.15  &  $6_{-1}^{+1}$          &    $0.01_{-0.01}^{+0.01}$& $-6.81_{-0.19}^{+0.19}$   \\    
    50  & 12 28 09.21    & 44 05 40.42 &  0.04 (0.7) &  0.27   &  20.99$\pm$0.01   &  20.83$\pm$0.01  &  20.45$\pm$0.01  &   -7.17$\pm$0.15  &  $120_{-9}^{+16}$          &    $0.10_{-0.01}^{+0.01}$& $-9.18_{-0.17}^{+0.16}$   \\    
   51$^{\dagger}$$^{\star}$   & 12 28 10.04    & 44 05 24.95 &  0.16 (3.0) &  0.09   &  20.46$\pm$0.01   &  19.60$\pm$0.01  &  19.72$\pm$0.01  &   -9.30$\pm$0.15  &  $5_{-1}^{+1}$          &    $0.02_{-0.01}^{+0.01}$& $-8.39_{-0.15}^{+0.15}$   \\    
   52$^{\ast}$(14)  &    12  28  06.64  & 44  06  07.78   &  0.13     (2.4) &  0.13   &  19.95$\pm$0.01   &  19.15$\pm$0.01  &  18.01$\pm$0.01 &   -8.91$\pm$0.15  &  $10013_{-1473}^{+1708}$   &  $11.79_{-1.06}^{+1.12}$  & $-14.51_{-0.19}^{+0.20}$   \\    
   53$^{\ast}$(34)  &    12  28  11.46  & 44  05  00.57   &  0.17     (3.1) &  0.11   &  20.00$\pm$0.01   &  19.78$\pm$0.01  &  19.33$\pm$0.03 &   -8.22$\pm$0.15  &  $ 251_{-14}^{+6}$         &   $0.42_{-0.02}^{+0.01}$ & $-10.75_{-0.15}^{+0.16}$   \\    
   54  & 12 28 08.82    & 44 05 33.34 &  0.12 (2.2) &  0.19   &  21.53$\pm$0.01   &  21.34$\pm$0.02  &  20.83$\pm$0.03  &   -6.67$\pm$0.15  &  $257_{-2}^{+2}$          &    $0.11_{-0.01}^{+0.01}$& $-9.21_{-0.15}^{+0.15}$   \\    
   55$^{\ast}$(23)$^{\star}$  &    12  28  09.74  & 44  05  19.71   &  0.07     (1.3) &  0.33   &  19.21$\pm$0.02   &  19.29$\pm$0.02  &  19.42$\pm$0.05 &   -8.67$\pm$0.15  &  $  6_{-1}^{+1}$           &   $0.05_{-0.01}^{+0.01}$ & $ -8.25_{-0.17}^{+0.18}$   \\    
   56$^{\dagger}$$^{\ast}$(22)$^{\star}$  &    12  28  09.65  & 44  05  16.27   &  0.04     (0.7) &  0.26   &  18.09$\pm$0.01   &  17.91$\pm$0.01  &  18.02$\pm$0.01 &  -10.06$\pm$0.15  &  $   7_{-1}^{+2}$          &   $0.13_{-0.01}^{+0.01}$ & $ -9.43_{-0.18}^{+0.18}$   \\    
    57$^{\ast}$(24)  &    12  28  09.90  & 44  05  09.83   &  0.09     (1.6) &  0.05   &  18.38$\pm$0.01   &  18.51$\pm$0.02  &  18.68$\pm$0.04 &   -9.45$\pm$0.15  &  $   5_{-1}^{+1}$          &   $0.12_{-0.01}^{+0.01}$ & $ -9.15_{-0.15}^{+0.15}$   \\    
   58$^{\ast}$(13)  &    12  28  06.44  & 44  05  55.12   &  0.08     (1.4) &  0.12   &  19.65$\pm$0.01   &  18.87$\pm$0.01  &  17.76$\pm$0.01 &   -9.18$\pm$0.15  &  $ 7760_{-670}^{+1550}$   &  $12.62_{-0.64}^{+1.57}$ & $-14.58_{-0.21}^{+0.17}$   \\    
    59  &    12  28  07.95  & 44  05  23.48   &  0.14     (2.6) &  0.11   &  21.23$\pm$0.03   &  21.00$\pm$0.03  &  20.46$\pm$0.06  &   -7.02$\pm$0.15  &  $ 270_{-4}^{+4}$          &   $0.14_{-0.01}^{+0.01}$ & $ -9.58_{-0.15}^{+0.15}$   \\    
    60  &    12  28  09.59  & 44  04  58.43   &  0.25     (4.6) &  0.14   &  18.97$\pm$0.01   &  18.94$\pm$0.01  &  18.67$\pm$0.02  &   -9.06$\pm$0.15  &  $   9_{-1}^{+1}$          &   $0.09_{-0.01}^{+0.01}$ & $ -8.91_{-0.15}^{+0.15}$   \\    
    61  &    12  28  08.17  & 44  05  11.01   &  0.13     (2.4) &  0.22   &  20.67$\pm$0.02   &  20.44$\pm$0.02  &  19.98$\pm$0.05  &   -7.57$\pm$0.15  &  $ 253_{-22}^{+10}$        &   $0.23_{-0.02}^{+0.01}$ & $-10.10_{-0.15}^{+0.17}$   \\    
   62  &    12  28  07.64  & 44  05  12.58   &  0.10     (1.9) &  0.13   &  22.70$\pm$0.04   &  22.25$\pm$0.09  &  21.20$\pm$0.09  &   -5.80$\pm$0.17  &  $1265_{-117}^{+115}$      &   $0.13_{-0.01}^{+0.01}$ & $ -9.44_{-0.17}^{+0.17}$   \\    
    63$^{\ast}$(17)  &    12  28  07.40  & 44  05  15.07   &  0.19     (3.6) &  0.09   &  20.97$\pm$0.03   &  20.61$\pm$0.03  &  19.97$\pm$0.06 &   -7.41$\pm$0.15  &  $ 537_{-167}^{+97}$      &   $0.30_{-0.06}^{+0.03}$ & $-10.48_{-0.19}^{+0.32}$   \\    
   64  & 12 28 10.24    & 44 04 36.28 &  0.05 (0.9) &  0.26   &  21.80$\pm$0.01   &  21.62$\pm$0.01  &  21.09$\pm$0.01  &   -6.39$\pm$0.15  &  $212_{-9}^{+5}$          &    $0.07_{-0.01}^{+0.01}$& $-8.77_{-0.15}^{+0.15}$   \\    
   65$^{\ast}$(19)  &    12  28  08.82  & 44  04  52.01   &  0.26     (4.9) &  0.16   &  19.87$\pm$0.02   &  19.61$\pm$0.02  &  19.07$\pm$0.03 &   -8.41$\pm$0.15  &  $ 275_{-7}^{+12}$        &   $0.52_{-0.01}^{+0.01}$ & $-10.98_{-0.15}^{+0.15}$   \\    
    66  &    12  28  07.86  & 44  05  03.49   &  0.38     (7.1) &  0.15   &  19.73$\pm$0.02   &  19.44$\pm$0.02  &  18.73$\pm$0.04  &   -8.58$\pm$0.15  &  $ 386_{-15}^{+20}$        &   $0.76_{-0.02}^{+0.02}$ & $-11.40_{-0.16}^{+0.15}$   \\    
    67  &    12  28  09.34  & 44  04  38.99   &  0.42     (7.8) &  0.11   &  19.33$\pm$0.01   &  19.05$\pm$0.02  &  18.43$\pm$0.03  &   -8.97$\pm$0.15  &  $ 307_{-8}^{+11}$        &   $0.92_{-0.01}^{+0.01}$ & $-11.60_{-0.15}^{+0.15}$   \\    
   68$^{\ast}$(11)  &    12  28  06.24  & 44  05  15.98   &  0.16     (2.9) &  0.03   &  21.11$\pm$0.03   &  20.89$\pm$0.03  &  20.39$\pm$0.07 &   -7.11$\pm$0.15  &  $ 261_{-3}^{+3}$          &   $0.16_{-0.01}^{+0.01}$ & $ -9.66_{-0.15}^{+0.15}$   \\
   69  &    12  28  07.02  & 44  04  51.82   &  0.24     (4.4) &  0.06   &  21.63$\pm$0.07   &  20.95$\pm$0.05  &  19.91$\pm$0.07  &   -7.10$\pm$0.16  &  $1428_{-1}^{+1}$          &   $0.35_{-0.01}^{+0.01}$ & $-10.63_{-0.16}^{+0.16}$   \\    
   70  & 12 28 04.22    & 44 05 27.95 &  0.07 (1.2) &  0.06   &  21.97$\pm$0.01   &  21.68$\pm$0.01  &  21.12$\pm$0.02  &   -6.33$\pm$0.15  &  $294_{-16}^{+17}$          &    $0.08_{-0.01}^{+0.01}$& $-8.94_{-0.15}^{+0.15}$   \\    
   71  &    12  28  03.09  & 44  05  31.92   &  0.32     (6.0) &  0.04   &  20.23$\pm$0.02   &  20.01$\pm$0.03  &  19.59$\pm$0.06  &   -8.00$\pm$0.15  &  $ 251_{-18}^{+5}$         &   $0.33_{-0.02}^{+0.01}$ & $-10.53_{-0.15}^{+0.17}$   \\    
   72$^{\ast}$(8)  &    12  28  04.69  & 44  04  53.42   &  0.12     (2.2) &  0.04   &  21.17$\pm$0.02   &  20.36$\pm$0.02  &  19.21$\pm$0.02 &   -7.70$\pm$0.15  &  $10560_{-1241}^{+2324}$        &   $4.02_{-0.28}^{+0.51}$      & $-13.34_{-0.22}^{+0.18} $   \\    
   73$^{\ast}$(15)  &    12  28  06.87  & 44  04  20.51   &  0.07     (1.4) &  0.12   &  19.49$\pm$0.01   &  19.62$\pm$0.01  &  19.80$\pm$0.02 &   -8.34$\pm$0.15  &  $   5_{-1}^{+1}$          &   $0.04_{-0.01}^{+0.01}$ & $-8.03_{-0.15}^{+0.15} $   \\    
   74$^{\ast}$(3)  &    12  28  02.54  & 44  05  18.39   &  0.23     (4.3) &  0.13   &  21.57$\pm$0.06   &  21.40$\pm$0.08  &  21.13$\pm$0.22 &   -6.60$\pm$0.17  &  $ 110_{-42}^{+3}$        &   $0.06_{-0.01}^{+0.01}$ & $-8.57_{-0.17}^{+0.34} $   \\    
   75$^{\ast}$(5)  &    12  28  02.61  & 44  04  51.51   &  0.26     (4.7) &  0.11   &  21.29$\pm$0.01   &  20.62$\pm$0.01  &  19.58$\pm$0.02 &   -7.43$\pm$0.15  &  $1427_{-1}^{+1}$          &   $0.47_{-0.01}^{+0.01}$ & $-10.96_{-0.15}^{+0.15}$   \\    
   76$^{\ast}$(7)  &    12  28  03.88  & 44  04  15.26   &  0.17     (3.2) &  0.03   &  20.51$\pm$0.01   &  19.70$\pm$0.01  &  18.52$\pm$0.01 &   -8.36$\pm$0.15  &  $12000_{-2575}^{}$        &   $8.19_{-1.08}^{}$      & $-14.10_{}^{+0.24}     $   \\    
   77  &    12  27  57.53  & 44  05  28.16   &  0.10     (1.9) &  0.24   &  19.29$\pm$0.01   &  18.51$\pm$0.01  &  17.42$\pm$0.01  &   -9.55$\pm$0.15  &  $7244_{-792}^{+737}$    &  $16.60_{-1.20}^{+1.12}$ & $-14.88_{-0.17}^{+0.18}$   \\    
   78$^{\dagger}$$^{\star}$  & 12 27 56.63     &44 05 29.32 &  0.15 (2.8) &  0.20   &  21.64$\pm$0.01   &  20.14$\pm$0.01  &  20.73$\pm$0.03  &   -7.78$\pm$0.15  &  --         &   -- & --  \\    
   79  &    12  27  59.87  & 44  04  43.34   &  0.10     (1.9) &  0.27   &  18.99$\pm$0.01   &  18.19$\pm$0.01  &  17.09$\pm$0.01  &   -9.87$\pm$0.15  &  $8130_{-624}^{+573}$      &   $24.19_{-1.29}^{+0.98}$& $-15.30_{-0.16}^{+0.16}$   \\    
   80  &    12  27  57.66  & 44  04  44.91   &  0.15     (2.8) &  0.04   &  20.77$\pm$0.01   &  20.01$\pm$0.01  &  18.90$\pm$0.01  &   -8.05$\pm$0.15  &  $1446_{-3}^{+4}$         &    $0.75_{-0.01}^{+0.01}$& $-11.49_{-0.15}^{+0.15}$   \\    
   81  &    12  27  52.92  & 44  04  53.42   &  0.11     (2.0) &  0.14   &  21.08$\pm$0.01   &  20.42$\pm$0.01  &  19.44$\pm$0.01  &   -7.63$\pm$0.15  &  $1418_{-1}^{+2}$          &    $0.59_{-0.01}^{+0.01}$& $-11.21_{-0.15}^{+0.15}$   \\    
\enddata
\tablenotetext{a}{The $\ast$ indicates clusters in common with \cite{gel01}. For these clusters we provide also the ID assigned by \cite{gel01} in Table~3. The $\dagger$ indicates clusters with $B-V>0$, $V-I<0$ 
(see Section~\ref{cluster_agemass}). The $\star$ indicates clusters associated with H$\alpha$ emission as shown in Fig.~\ref{clu_ha}.}
\tablenotetext{b}{Intrinsic effective radius in arcsec and parsec (adopting a distance of 3.82 Mpc from A08) derived with ISHAPE (see Section~\ref{clu_size_eps}).}
\tablenotetext{c}{Cluster ellipticity  computed as $1 - \eta$, where $\eta$ is the ISHAPE minor over major axis ratio (see Section~\ref{clu_size_eps}).}
\tablenotetext{d}{Cluster total magnitudes in the ACS Vegamag system (see Section~\ref{photometry}).}
\tablenotetext{e}{Absolute Johnson-Cousins V cluster magnitude computed adopting a distance of 3.82 Mpc from A08 and correcting for a Galactic reddening 
of $E(B-V)=0.019$ (see Section~\ref{photometry}). 
A reddening correction of $E(B-V)=0.3$ was applied to cluster 51 (see Section~\ref{cluster_agemass}).}
\tablenotetext{f}{Cluster age and mass derived with the Z$=$0.004 Padova models \citep{pad10} and assuming a Salpeters' IMF down to 0.1 $M_{\odot}$. Clusters with $\dagger$ were fitted using the   Z$=$0.004 GALEV models
(see Section~\ref{cluster_agemass}). Only Galactic reddening was assumed, with the exception of cluster 51, for which we solved for reddening and derived $E(B-V)\sim0.3$.}
\tablenotetext{g}{Cluster absolute V magnitude at 10 Myr obtained with the  Z$=$0.004 Padova models (see Section~\ref{cluster_classification}), and with the  Z$=$0.004 GALEV models for $\dagger$ .}
\end{deluxetable}

\subsection{Sizes and ellipticities\label{clu_size_eps}}

Intrinsic sizes of the candidate star clusters were derived using 
the ISHAPE task in BAOLAB \citep{lar99}. 
ISHAPE models a source as an analytical function convolved with the PSF of the image. 
For each object, ISHAPE starts from an initial value for the FWHM, ellipticity, 
amplitude, and object position. 
These parameters are then adjusted 
in a $\chi^2$ iterative minimization, until the 
best fit between the observed profiles and the model 
convolved with the PSF is obtained. We adopted a fitting radius 
of $\approx$ 6 $\times$ FWHM$_{PSF}$ in the ISHAPE procedure, 
corresponding to $\sim$ 0.5'' ($\sim$ 14 pixels). 
We implemented both a King (1962) profile 
with a concentration parameter c $=$ 30  (where 
c is the ratio between the tidal radius and the core radius), 
and a Moffat profile with a power index of 1.5.
Both profiles provide reasonable descriptions of the brightness profiles of GCs 
 \citep[see][for an extensive discussion of GC brightness profiles]{mclau05}.
The ISHAPE output includes the
intrinsic major axis FWHM, minor/major axis ratio, $\chi^2$, flux, and signal-to-noise ratio for
each object, plus a residual image. 
Intrinsic effective radii $R_e$ were obtained multiplying the intrinsic FWHM,
averaged over the major and minor axes,  
times the conversion factors (1.48 for {\it KING30} and 1.13 for {\it MOFFAT15}) provided in Table~3 of the ISHAPE tutorial.
King and Moffat profiles do not provide significant 
difference in the minimum $\chi^2$ or in the final intrinsic $R_e$.
We provide in  Col.~4 of Table~1 the $R_e$ averaged over the F435W, F555W, 
and F814W images, both in arcsec and in parsec adopting a distance of 3.82 Mpc from A08.
For some clusters a best fit could not be found in one of the three bands. 
In this cases, the final $R_e$ was obtained by averaging only in two bands.
 
Ellipticities were derived as $\epsilon = 1 - \eta$, where $\eta$ is the ISHAPE minor over major axis ratio averaged over the three bands.
The values for the individual clusters are provided in Col.~5 of Table~1.
The $R_e$ and $\epsilon$ distributions for blue ($m_{F555W}-m_{F814W}<0.9$ or $m_{F435W}-m_{F555W}<0.5$) and red  
($m_{F555W}-m_{F814W}\ge0.9$ and $m_{F435W}-m_{F555W}\ge0.5$) clusters (see Section~\ref{photometry} and Table 1) are shown in Fig.~\ref{reffeps}.
$R_e$ varies from $\sim$ 1 pc to $\sim$ 9 pc, with a median value of $\sim$ 3.2 pc, while the ellipticity ranges from 
$\epsilon \sim 0$ to 0.4, with a median value of $\sim$ 0.1, and with the most elliptical clusters ($\epsilon>0.3$) being all blue.

\subsection{Photometry \label{photometry}}

Aperture photometry with the PHOT package within IRAF was performed at the cluster positions within a radius aperture 
$R_{phot}=\sqrt{FWHM_{M15}^2 + FWHM_{PSF}^2}$, where $FWHM_{M15}$ is the ISHAPE intrinsic cluster size 
obtained with a {\it MOFFAT15} profile, averaged over the major and minor axes and over the three 
photometric bands, and $FWHM_{PSF}\sim$ 2.6 pixels ($\sim$0.1'').
The sky was computed in a 5 pixel annulus at  5$\times R_{phot}$.
Aperture corrections from $R_{phot}$ to ``infinity'' were computed  
from the results of the artificial cluster experiments described in Section~\ref{artificial}, and applied to 
the photometry to obtain total magnitudes. The artificial cluster tests also provide  an estimate of 
the photometric error as a function of magnitude, cluster size, and photometric band. 
Finally, the photometry was calibrated into the HST Vegamag system adding the zeropoints provided in \cite{sir05}.
The total F435W, F555W, and F814W magnitudes are given in Cols.~6, 7, and 8 of Table~1. In Col.~9 we 
provide the absolute V magnitudes in Johnson-Cousins obtained with the transformations 
of  \cite{sir05}, corrected for a Galactic-extinction of E(B-V)$=$0.019 \citep{sch98}, and adopting a distance modulus of 27.91$\pm$0.15 
($3.82\pm0.27$ Mpc, A08). 
In Fig.~\ref{gelattus} we compare the observed V magnitudes  with those measured by \cite{gel01} for the 37 clusters in common.
The agreement is pretty good, apart from some discrepancy at the brightest magnitudes. 
Bright clusters tend to be located in crowded regions of high star formation, and their photometry is more uncertain. 
In particular, there may be some differences in the adopted photometry aperture between us and \cite{gel01}, since their
photometry extends ``up to the radius where the clusters visually merged into the background''.

\subsection{Artificial cluster experiments \label{artificial}}

We ran experiments with artificial clusters to evaluate the level 
of completeness and photometric error for different magnitudes, cluster sizes,  and crowding conditions.
These tests also provide aperture corrections to the cluster photometry from the adopted 
$R_{phot}$ aperture to ``infinity''.

First, we generated models of extended sources using the Baolab task 
MKCMPPSF \citep{lar99}. This task creates a model source 
by convolving a user-supplied PSF (in this case the ACS PSF created from our images) with 
an analytic profile (we chose a {\it MOFFAT15} profile) with variable FWHM.
We simulated model sources with intrinsic FWHM of 2, 3, 4, 5, 7, 9, 11 and 13 pixels.
 
Then, using the MKSYNTH task within Baolab, we generated 
(2000$\times$2000) pixel$^2$ zero-background images  
with the model sources homogeneously distributed in a 10$\times$10 array.
In this way the artificial cluster separation is 200 pixels.
In MKSYNTH, each source is built up by assigning 
to each photon a random position with probability determined by 
the PSF and the intrinsic source profile.
The number of electrons $N_e$ added for an object of magnitude $mag$ is 
$N_e=MKSYNTH.ZPOINT \times 10^{-0.4 \times mag}$, and thus 
$MKSYNTH.ZPOINT=t_{exp} \times 10^{0.4 \times ZP}$, where 
$t_{exp}$ is the total exposure time and $ZP$ is the photometric zero point 
in the specific photometric band.
The procedure was repeated for all the possible combinations of size (FWHM$=$2, 3, 4, 5, 7, 9, 11, 13 pixels) and 
magnitudes ($m_{HST,Vega}$ from 17 to 22.5 mag, with a step of 0.5),     
resulting in 12$\times$8 = 96 images per filter.
Each synthetic image of given (FWHM, $m_{Vega}$) was
added to the real image in one of the three galaxy subregions 
shown in Figs.~\ref{image_compl} using the IMARITH task within IRAF,  and 
the data reduction process was repeated in the same way as for the real stars (see A08) 
in a (2000$\times$2000) pixel$^2$ portion of the original science image. 
The three selected fields account for the different crowding conditions within the NGC~4449 field of view, 
with Field 1, chosen on the galaxy center, being the most crowded one, and Field 3, selected on 
a low-background region far from the galaxy center, being the least crowded one.
Clusters were selected from the final photometric catalog, calibrated as the real star catalog, adopting a sharpness cut 
of $sharpness_{DAO} \geq$ 0.5 and, for the F814W images, a magnitude cut of $m_{F814W} < 22$. 
The sharpness/magnitude selected catalog was then cross-correlated with the 
coordinate list given in input to  MKSYNTH to generate the synthetic images.
The ISHAPE task was run on the matched catalog to determine the size of the artificial clusters.
Aperture photometry with PHOT was then performed within an aperture radius $R_{phot}$ in the same way 
as for the real clusters (see Section~\ref{photometry}). 
For each filer, this process was repeated 12 (mags) $\times $ 8 (FWHMs) $\times$ 3 (fields) times, 
providing 28,800 simulated clusters in each band.
The results of the artificial cluster experiments are summarized in Fig.~\ref{compl} and \ref{errors}.
In Fig.~\ref{compl}, we show the fraction of recovered clusters as a function of the F555W magnitude for 
 clusters of different sizes in the three selected fields.
 The completeness is close to 100\% at the brightest magnitudes, and decreases 
 with increasing magnitude. At a fixed magnitude, the completeness is lower for more extended 
 objects and for more crowded fields, as already found by \cite{mora07} and \cite{smith07}.  

 We also notice that stochastic effects induced by the IMF sampling, not included in our study, 
 can affect the cluster detection rate; however, this effect starts to be significant for clusters with masses below $10^4 M_{\odot}$ 
 \citep{silva11}, while the majority of our detected clusters have masses above this limit (see Section \ref{cluster_agemass}). 

Fig.~\ref{errors} shows the input minus output F555W magnitude versus the input magnitude for the recovered artificial clusters in the three 
fields. In this example, the clusters have an intrinsic FWHM$=$3 pixel ($\sim$0.1'').
The input magnitude is the total cluster magnitude, while the output magnitude results from 
the photometry within an $R_{phot}$ radius aperture (see Section~\ref{photometry} for details).
Therefore, the average mag(inp)$-$mag(out) is the aperture correction from  $R_{phot}$ to ``infinite''
to be applied to the real cluster photometry.
We notice in Fig.~\ref{errors} that the aperture correction is the same in all the three fields at the brightest magnitudes.
Then it decreases as magnitude increases, and this effect is larger as the field is more crowded 
($\sim$0.2 mag difference at F555W$=$22.5 in Field 1).
This trend is due to the increasing importance at fainter cluster magnitudes of the contamination from other sources 
falling within $R_{phot}$.
We determined for each real cluster the ``most representative field'' by computing the average surface brightness 
in a (30-100) pixel annulus from the cluster center, and comparing this value with the average surface brightness of the three fields.  
Then, given the field, the band, and the cluster $R_{phot}$ magnitude, we applied the appropriate correction to ``infinity'',
which accounts also for the effect of blending with other sources.
The 1 $\sigma$ dispersion level around the average mag(inp)$-$mag(out) value was adopted as photometric error.
 
The artificial cluster experiments provide also a test on the
reliability of the cluster size measurements.
We show in Fig.~\ref{fwhmtest} the average input minus output FWHM for the artificial clusters as a function of the 
F555W magnitude, for different FWHM input values and for the three test fields.
Fig.~\ref{fwhmtest} reveals a trend toward recovering larger cluster sizes, with the differences in  FWHM 
being larger at fainter magnitudes and for more extended objects, in agreement with \cite{mora07}. The differences tend to increase 
from Field 3 to Field 1, i.e. going from low-background regions to high-background, crowded regions.
In Field 2, we derive an average FWHM(inp)$-$FWHM(out)$\sim-0.06$ pixel ($\sim$ 1\% difference) for a cluster with a magnitude of F555W$=$20 and a size of FWHM$=$5 pixel, while the difference is as large as  $\sim -0.3$ pixels ($\sim$6 \%) at  F555W$=$22.5.
 
\section{Results \label{resu}}
 
\subsection{Luminosity Functions \label{cluster_lf}}

We show in Fig.~\ref{lf} the observed and completeness-corrected cluster luminosity functions (LFs) in F435W, F555W, and F814W.
The completeness-corrected LFs were obtained  ``replicating'' $1/f_{compl}$ times each observed cluster of given 
position in the NGC~4449 field of view, size, and 
($m_{F435W}$, $m_{F555W}$, $m_{F814W}$) magnitudes, where $f_{compl}$ is  the 
product of the completeness in the three bands for that cluster size and crowding conditions (e.g. Field 1, 2 or 3, see 
Section~\ref{artificial}), and varies from 0 to 1.

The histograms in the top row of panels of Fig.~\ref{lf} refer to the total sample. 
We are almost 100\% complete at magnitudes brighter than 
$m_{F435W}, m_{F555W}\sim 20$ and $m_{F814W}\sim19$, while at fainter 
magnitudes we are strongly incomplete, and the correction to the LF is very large.
At magnitudes fainter than $\sim$22, where our completeness drops to zero, we are not able to recover the LFs.

Particular attention must be payed when deriving the slope of the LF, because the fact that the magnitude detection limit of the clusters shifts toward brighter magnitudes with increasing clusters size (see Fig.~\ref{compl}) can create ``artificial'' slopes. 
At the faintest magnitude bins, where only the most compact clusters are detected, we are not able to recover the contribution to the LF from the extended clusters.
For this reason we restrict our analysis to clusters with $R_e\le4$ pc and to magnitudes brighter than $\sim$21, which is 
the detection limit for clusters of size $R_e\sim4$ pc. For magnitudes fainter than this we are not able to properly correct the LFs.
In the middle and bottom panels of Figs.~\ref{lf}, 
the size-selected LFs are shown separately for blue  ($m_{F555W}-m_{F814W}<0.9$ or $m_{F435W}-m_{F555W}<0.5$) 
and red ($m_{F555W}-m_{F814W}\ge0.9$ and $m_{F435W}-m_{F555W}\ge0.5$) clusters, 
roughly corresponding to young clusters and old globular clusters (see Section~\ref{cluster_agemass}).
We derived  linear least squares fits to the histograms in the form $\log N= a m + b$. 
Then, from the slope $a$, we derived the slope $\alpha$ of the luminosity function $dN(L)\propto L^{\alpha} dL$ using 
$\alpha=-(2.5a + 1)$. The values of $\alpha$ derived for the blue and red cluster populations are provided in 
Fig.~\ref{lf} and in Cols.~2 and 5 of Table~\ref{slopes}.  For the young clusters, the slopes $\alpha$ are flatter than $\sim-1.5$. 

These slopes should be taken with caution. In fact, it has been shown that in the low 
number regime, direct fitting to binned histograms can lead to systematic errors in the derived indices 
\citep[e.g.][]{maiz05}. One way around this problem  is to use a maximum likelihood estimator (MLE), as shown 
by \cite{masch09}. Following this approach, and using the same blue and red cluster samples as  
for the binned histogram fitting (BHF) method, we derived new slopes, which we provide in Cols. 3 and 6 of Table~\ref{slopes}. 
The BHF and MLE approaches provide consistent results, within the errors. 
In Table~\ref{slopes} we also report the magnitude interval adopted to derive the slopes in each band and for the blue and red cluster samples, respectively.

It is interesting to compare our slopes with those derived in the literature for other galaxies. 
In Fig.~\ref{alpha} we present an updated version of the diagram proposed by \cite{gieles10} which shows the results of power law fits to LFs of clusters in different 
galaxies taken from the literature.  In the V band, the slopes fall in the range $-2.7\lesssim\alpha\lesssim-1.4$. 
Spiral galaxies exhibit slopes typically steeper than $\alpha\sim-2$. If we exclude blue compact galaxies (BCGs), there is a trend in the sense that the LF becomes steeper with increasing luminosity \citep[see][for a discussion]{gieles10}. BCGs deviate from this trend exhibiting flat slopes at bright luminosities. 
Indeed it has been noticed that the cluster LF in BCGs tend to be flatter than the typical values  found in spirals 
 \citep{ostlin03,adamo10,adamo11,adamo11b}. A possible explanation is that the environment in these  starburst galaxies has favored  the formation 
 of massive clusters and/or the disruption of low mass clusters, even if  blending can flatten the observed LF \citep{adamo11}.  
 The slopes derived in NGC~4449 tend to be flatter than the values found in spirals, and more similar to those of BCGs; 
 its location in the $\alpha$ versus luminosity diagram appears  intermediate between that of spirals and BCGs.


\begin{deluxetable}{ccccccc}
\tabletypesize{\scriptsize}
\tablecolumns{7}
\tablecaption{Luminosity function slopes}
\tablehead{
\colhead{} & \multicolumn{3}{c}{Blue clusters}\tablenotemark{a} &  \multicolumn{3}{c}{Red clusters}\tablenotemark{b} \\
\colhead{Band} & \colhead{$\alpha$ (BHF)}\tablenotemark{c}   & \colhead{$\alpha$ (MLE)}\tablenotemark{d}   
& \colhead{Fit interval}  &  \colhead{$\alpha$ (BHF)}\tablenotemark{c}  
& \colhead{$\alpha$ (MLE)}\tablenotemark{d}  & \colhead{Fit interval} \\     
} 
\startdata
F435W & $-1.37\pm0.30$  & $-1.36$ &  $-11.0<M_{F435W}< -7.0$  &    $-1.60\pm0.35$  & $-1.37$  &  $-9.5<M_{F435W}< -7.0$ \\
F555W & $-1.47\pm0.22$  & $-1.45$ &  $-11.0<M_{F555W}< -7.0$  &    $-1.30\pm0.20$  & $-1.15$  &  $-10.0<M_{F555W}< -7.5$ \\
F814W & $-1.47\pm0.12$  & $-1.52$ &  $-11.5<M_{F814W}< -7.5$  &    $-1.10\pm0.22$  & $-1.07$  &  $-11.0<M_{F814W}< -8.5$ \\
\enddata
\tablenotetext{a}{$m_{F555W}-m_{F814W}<0.9$ or $m_{F435W}-m_{F555W}<0.5$.}
\tablenotetext{b}{$m_{F555W}-m_{F814W}\ge0.9$ and $m_{F435W}-m_{F555W}\ge0.5$.}
\tablenotetext{c}{Binned histogram fitting.}
\tablenotetext{d}{Maximum likelihood estimator.}
\label{slopes}
\end{deluxetable}

\subsection{Ages and Masses \label{cluster_agemass}}

The F435W$-$F555W (B$-$V) versus F555W$-$F814W (V$-$I) color-color diagram 
for the candidate star clusters is shown in Fig.~\ref{colcol}. 
We can distinguish four different populations in the diagram: 
red clusters (likely old globular clusters) with $V-I>0.9$ and  $B-V>0.5$; 
blue (and young) clusters with $0<V-I<0.9$ and $0<B-V<0.5$; very 
blue (and very young) clusters with $V-I<0$ and $B-V<0$; V-bright clusters, with $V-I<0$ and $B-V>0$ 
\citep[likely young clusters embedded in ionized gas, see e.g.][]{fedotov11}.
We count 4 very blue clusters (clusters number 13, 55, 57, and 73), 6 V-bright clusters (clusters number 10, 12, 49, 51, 56, 78), 
44 blue clusters, and 27 red clusters. In the figure, we also indicate the 11 clusters of Fig.~\ref{clu_ha} associated with H$\alpha$ emission. Nebular emission is present in all the 6 V-bright clusters, in 2 of the very blue clusters, and in 3 of the blue clusters.

We overplot the Padova-2010 simple stellar population (SSP) models{\footnote{downloaded at {\tt http  //stev.oapd.inaf.it/cgi-bin/cmd}}}  \citep{pad10} and the GALEV models \citep{galev} for a \cite{salp} initial mass function (IMF), a Galactic reddening 
E(B-V)=0.019 \citep{sch98} and metallicities from Z$=$0.0004 to Z$=$0.02. 
The Padova SSPs are based on the \cite{marigo08} models 
(e.g., the \cite{gir00} tracks up to the early-AGB $+$ the detailed TP-AGB modeling from 
\cite{mg07} for $M\le 7 M_{\odot}$ $+$ the \cite{bert94} models for $M>7 M_{\odot}$)
with the \cite{pad10} correction for low-mass, low-metallicity AGB tracks.
The spectral library and bolometric corrections are described in \cite{girardi08} and \cite{loidl01}.
The GALEV models are based on the isochrones of  \cite{gir00} and on the 
model atmosphere spectra from \cite{leje97,leje98}, and include gaseous 
continuum and line emission from ionized gas. The effect of the nebular emission 
on the integrated colors is visible as the turn-over of the models at the youngest ages.

We notice that the Padova models account for the red, blue, and very blue cluster populations, 
but can not account for the 6 clusters with $V-I<0$, $B-V>0$. If we exclude cluster 56,
in all the other 5 clusters the nebular emission is spatially coincident with the cluster stars (see Fig.~\ref{clu_ha}), 
and thus the integrated colors are highly contaminated by the ionized gas.  
This is why the GALEV models, which include a treatment of the nebular emission, 
provide a better match than Padova's. According to the H$\alpha$-morphology classification proposed by \cite{whit11} these are {\it emerging} clusters (category 3) with a mean age of $\sim$3 Myr. On the other hand, the GALEV models 
fail in reproducing the very blue clusters with $V-I<0$ and $B-V<0$, for which the Padova models provide a good match and 
indicate ages of a few Myr. Two (13, 55) of these clusters are shown in Fig.~\ref{clu_ha}: the H$\alpha$ emission is not coincident 
with the cluster itself, but has the form of an envelope surrounding the cluster, likely pushed out by massive star winds and supernovae.
Thus the cluster integrated colors, measured on a relatively small aperture, are not affected by the ionized gas contribution, and are well reproduced by the Padova models. According to \cite{whit11}, these clusters belong to category 4a, with a mean age of $\sim$5 Myr.

Models of different metallicities are strongly degenerate in the (B$-$V) versus (V$-$I) 
diagram. In fact, while we can safely exclude a metallicity as low as  Z$=$0.0004, which fails 
to reproduce the red cluster population, models of higher metallicity 
(Z$=$0.004, 0.008, and 0.02) are all consistent with the cluster colors.
We notice in particular that the Z$=$0.02 models are those that 
best reproduce the blue cluster population, and this holds in particular for the GALEV models. 
This is surprising, 
because abundance estimates in NGC~4449 HII regions
 \citep{talent,hgr82,mar97} give an average value 
 of 12 $+$ log(O/H)$_{N4449} \approx 8.31$; 
adopting the solar abundance determinations by \cite{caffau08,caffau09}
of  12 $+$ log(O/H)$_{\odot} \approx 8.76$ and $Z_{\odot} \approx 0.0156$, 
and assuming that the chemical composition of NGC~4449 follows the 
solar partition, this translates into $Z_{N4449}=0.0055$. 
Furthermore, a metallicity of Z$\sim$0.004  is consistent 
with the colors of the bulk of the red giant branch (RGB) in NGC~4449 (see A08).
Thus, the models that seem to best reproduce the blue cluster populations 
have a metallicity that is $\sim$ 3.6 times higher than that of the HII regions and of the RGB stars 
in NGC~4449. Similarly, \cite{gel01} noticed that the ensemble of clusters in NGC~4449 are 
fitted better with cluster evolutionary tracks of somewhat higher metallicity than one would have predicted 
from the nebular oxygen abundance. 

Other sources of uncertainties are the extinction toward the individual clusters 
and the IMF. 
The reddening vector shown in the panels of  Fig.~\ref{colcol} runs parallel to the age$\lesssim$1 
Gyr cluster sequence (for the GALEV models this does not hold for ages younger than $\sim$10 Myr), 
implying that an age-extinction degeneracy is present for the blue clusters.
However, the bluest clusters with $B-V<0$, $V-I<0$, which fall in proximity of the bluest Padova models, 
are not compatible with a high internal extinction. For the GALEV models, we show the effect of reddening 
in the left panel of  Fig.~\ref{colcol2}. The blue clusters are consistent with ages  younger than $\sim$100 Myr 
if the reddening is sufficiently high (up to $E(B-V)\sim0.2$).  The colors of the red clusters ($V-I>0.9$, $B-V>0.5$) 
imply old ages ($>$1 Gyr) and negligible internal extinction. Two of the $V-I<0$, $B-V>0$ clusters are compatible with 
negligible extinction (49, 56), one implies $E(B-V)\sim0.3$ (51), while three (10, 12, 78)  are too blue in $V-I$ for their $(B-V)$ 
at any reddening. It is actually possible that these are not clusters but background emission-line galaxies.

The effect of the IMF is shown in the right panel of Fig.~\ref{colcol2} for the 
GALEV models, where we plotted,  for a  Z$=$0.004 metallicity, Salpeter' s, Kroupa' s \citep {kroupa}, 
and Scalo' s \citep{scalo} IMFs. 
A Salpeter and Kroupa's IMF provide similar integrated colors, while 
some difference is observed at the youngest ages for a Scalo IMF. 
In general, the effect of the IMF on the integrated colors is very small 
at ages older than $\sim$ 1 Gyr.

For the three metallicities Z$=$0.004, 0.008, and 0.02, we derived the cluster ages minimizing 
the difference between models and data colors (B$-$V, V$-$I, and  B$-$I) 
according to a $\chi^2$ criterion, varying the age between $\sim$4 
Myr and $\sim$13 Gyr. The model colors were corrected for 
a Galactic reddening of E(B-V)$=$0.019. We did not solve for possible 
internal extinction, because of the small number of bands (3) and the 
relatively short wavelength range covered by our data. 
Following the prescriptions of \cite{avni}, 
the 1 $\sigma$ errors were derived as the limits of the 0.68 confidence level 
region defined by $\chi^2 < \chi^2_{min} + 1$, 
where $\chi^2$ is the actual value and not the reduced one.
Cluster masses were obtained from the mass-to-light ratios of the 
best-fitting models in the I band\footnote{Notice that an IMF flatter than Salpeter's below $\sim1M_{\odot}$, as suggested by different studies \citep[see e.g. Fig.2 in][]{bastian10}, results in cluster masses $\sim$30\%-40\% lower.}. 
The procedure was applied using both the GALEV models and the Padova models.
When using the Padova models, we did not fit the $V-I<0$, $B-V>0$ clusters.
For these clusters, the best fitting solution was searched with the GALEV models solving for reddening too, since 
at $V-I<0$ the GALEV models allow for a disentanglement of 
the age-reddening degeneracy. For clusters 10, 12, 78, with $B-V>1.3$, a satisfactory fit could not be obtained.

The results are shown in Fig.~\ref{agemass} in a mass versus age plot 
for the three assumed metallicities.
We have adopted a Salpeter IMF between 0.1 and 120 M$_{\odot}$ 
and a Galactic extinction  E(B-V)$=$0.019.
The age solutions obtained with the Z$=$0.004 models span the whole Hubble time, 
and the masses are in the range  $\sim10^3 - 2\times10^6 M_{\odot}$. 
We notice particular concentrations of clusters in correspondence of some age bins,
 such as at $\sim$ 6 Myr, 250 Myr, and 1.6 Gyr.  
 However, as discussed in \cite{bik03}, these peaks are not physical, and 
 result from the difficulty in deriving a good fit to the data in those regions were
 the model colors change very rapidly with time. 
Larger metallicities tend to provide younger ages, because of 
the age-metallicity degeneracy, and lower cluster masses. 
The Z$=$0.02 models provide a bimodal age distribution peaked at 
$\sim$ 6 Myr and $\sim$1.5 Gyr. 
At all metallicities, the GALEV models tend to produce younger ages 
than the Padova models.
 
In Fig.~\ref{agemass} we see a trend in the sense that the cluster mass 
tends to be higher with the cluster age. 
This is caused by a combination of effects. The lower envelope of the distribution 
is a completeness effect: because of the fading of the stellar populations with age,
old low-mass clusters are not detected. 
 To better show this effect we plot in Fig.~\ref{agemass} the line of a cluster  
of magnitude $m_{F555W}=21.5$ at different ages at the distance of NGC~4449. 
The upper envelope in the log(mass) versus log(age) distribution is instead due to a sampling effect.
In fact, in a log scale, bins of equal size sample age intervals which increase
with age. The probability of sampling a high mass cluster decreases with decreasing 
age interval, and this is why we do not observe high mass clusters at the youngest 
ages. See however Section~\ref{cluster_disruption} for more details. 
We derive a total mass in clusters of $\sim$1.7, $\sim$1.1, and $\sim$0.9 $\times$ 10$^7$ $M_{\odot}$ assuming 
metallicities of Z$=$0.004, 0.008, and 0.02, respectively (Padova models). 
These values are lower limits since low-mass old clusters are lost due to incompleteness. 

We provide in Cols.~10 and 11 of Table~1 the cluster ages and masses derived 
with the  Z$=$0.004 Padova models, 
since this metallicity is the closest to the nebular abundance and to the metallicity of the RGB stars in NGC~4449, even though 
we are not able to discriminate between the different metallicities from the integrated cluster colors alone.
For the $V-I<0$, $B-V>0$ clusters, the solution, when provided, was obtained with the GALEV models.
As it can be seen in Table~1, the ages of the clusters with associated H$\alpha$ emission, as derived from the 
integrated B V, I colors,  tend to fall in the range $\sim$ 5-7 Myr (clusters 13, 49, 51, 55, 56); clusters 9 and 43 have ages of $\sim$ 100 Myr, while for cluster 42  we derive an age as old as 
$\sim$ 300 Myr. We notice in Fig.~\ref{clu_ha} that for the youngest clusters the H$\alpha$ emission is closely associated with the cluster itself and has the form of an envelope surrounding it, while in the oldest ones the nebular emission is more extended and/or distant from the cluster, likely because it was pushed out by massive star winds and supernovae. These findings are in agreement with the results of \cite{whit11} pointing toward a close connection between the H$\alpha$ morphology/extent and the cluster age.

We analyze how clusters of different ages are distributed within the 
NGC~4449 field of view, and how they compare with stars in different age ranges 
as selected by A08. In Fig.~\ref{starclu} we show the spatial 
distribution of stars and clusters 
in four age bins ($<$10 Myr, 10$-$100 Myr, 100$-$1000 Myr, and $>$1Gyr).
For the clusters, the age selection is based on the results obtained assuming  
Z$=$0.004, for a self-consistent comparison with the selection of A08. 
There is a tendency for the distributions of stars and clusters to follow each-other. 
Young clusters are preferentially located in regions of young star formation. Old clusters are distributed 
over the whole NGC~4449 field of view, like the old stars. 
The difference between the spatial distribution of young clusters/stars and older stellar populations 
has also been noticed and quantified in the LMC \citep{bastian09b} and in other dwarf galaxies 
\citep[e.g.][]{annibali03,bastian11}. However, it is intriguing that in NGC~4449 the old clusters 
do not appear to be distributed homogeneously. As indicated in Fig.~\ref{starclu}, a number of 
linear structures are evident. It is possible that this distribution is a reflection of past satellite accretions in 
NGC~4449, possibly linked to its current ongoing starburst.
    
\cite{reines08} identified in NGC~4449 13 thermal radio sources, likely embedded massive clusters, and derived their physical 
properties from the nebular emission of the HII regions and from spectral energy distribution fitting to HST UV-near infrared data. The radio-detected clusters have ages 
$\lesssim$5 Myr and follow the distribution of the stars younger than 10 Myr in Fig.~\ref{starclu}.
Three of their thermal radio sources are in common with our cluster list, namely cluster 12, 51, and 55, which are displayed in Fig.~\ref{clu_ha}. For clusters 51 and 55 we derive ages of $5\pm1$ Myr and $6\pm1$ Myr, respectively. 

\subsection{Derivation of $\Gamma$}

Several authors have tried to estimate the total mass of clusters recently formed, and to 
compare this mass with the star formation rate over the same time scale. The ratio of stars forming in clusters is referred to 
as $\Gamma$ \citep{bastian08}. For ages younger than $\sim$10 Myr and masses above $\sim10^4M_{\odot}$, our cluster sample 
is likely to be complete (excluding young embedded clusters). Thus we computed the total mass in clusters more massive than this limit and younger than 
$\sim$10 Myr, and extrapolated it down to $10^3 M_{\odot}$ assuming a cluster initial mass function with slope $\alpha=-2$ \citep[a similar approach was 
adopted by, e.g.,][]{adamo11b}, to obtain the total mass in recently formed clusters. This mass was then divided by 5 Myr to obtain the recent cluster formation rate (CFR); younger clusters are likely to be lost in our images because still embedded or highly obscured. Considering all the results obtained with the Padova and GALEV models for different metallicities, 
we obtain $CFR=1.2\pm1.9\times10^{-2} M_{\odot} yr^{-1}$ which provides, adopting a current SFR of $\sim 1 M_{\odot} yr^{-1}$ from \cite{mcq10},
$\Gamma\sim0.01\pm0.02$. \cite{goddard10} found a positive correlation between $\Gamma$ and the star formation rate density $\Sigma_{SFR}$ for a sample 
including spiral and dwarf irregular galaxies. In Fig.~\ref{gamma} we present an updated version of the Goddard et al. diagram in which we include newer results 
from the literature, as well as the values derived for NGC~4449.  Here,  $\Sigma_{SFR}\sim0.04  M_{\odot} yr^{-1} kpc^{-2}$ was obtained dividing the current SFR by the total field of view of the two ACS pointings. Fig.~\ref{gamma} shows that NGC~4449 lies below the Goddard relation, exhibiting a too low CFR for its current SFR. However, if we include the central SSC, with an age between 6$-$10 Myr and a lower limit for the mass of $4\times10^5 M_{\odot}$ \citep{bok01}, $\Gamma$ becomes $\ga$0.09, 
and NGC~4449 is in better agreement with the relation. 
Another way to compute the CFR in NGC~4449 is to adopt the masses derived by \cite{reines08} for radio-detected embedded clusters, with ages $\lesssim$5 Myr.
From the masses listed in their Table 5 and 
adopting an age range of $\sim$5 Myr, we obtain $CFR=(4.3\pm0.5)\times10^{-2} M_{\odot} yr^{-1}$ and  $\Gamma\sim0.04\pm0.01$, higher than 
the previous value without the central SSC, but still below the Goddard relation.

\subsection{Cluster classification \label{cluster_classification}}

The terms of ``super star cluster'', ``populous cluster'', or ``massive cluster'' are not univocally defined in the literature. 
\cite{lari00} defined young massive clusters (YMCs) to be those with $(B-V)<0.45$ and $M_V <-8.5$, for $(U-B)>-0.4$, or $M_V <-9.5$, for $(U-B)<-0.4$. 
We do not have ACS imaging of NGC~4449 in the U band, but from the SSP models we see that  a $(U-B)=-0.4$ roughly corresponds to 
$m_{F435W}-m_{F555W} \sim 0.09$, accounting also for the Galactic reddening toward NGC~4449. 
Thus we can select YMCs to be those with ($0.09<m_{F435W}-m_{F555W} <0.45$, $M_V<-8.5$), and ($m_{F435W}-m_{F555W} <0.09$, $M_V<-9.5$). We end up with 12 YMCs, plus the central SSC. This provides a cluster specific frequency  of $T_N\sim0.68$\footnote{The young cluster specific frequency was defined by \cite{lari99} to be $T_N=N\times10^{0.4\times(M_B + 15)}$, where N is the number of YMCs, and $M_B$ is the total absolute B magnitude of the galaxy.}, given $M_{B}(NGC~4449)=-18.2$.
For their sample of 21 nearby spiral galaxies, \cite{lari00} derived $T_N$ in the range 0.00$-$1.77, with an average value of $\sim$0.60 and a median of 0.44. Thus, the specific frequency of YMCs in NGC~4449 is above the median and average values derived in spirals, 
likely because of its high SF activity, but not as extreme as the frequency derived in the most rich spirals.  
For comparison, the LMC has $T_N=0.57$ \citep{lari00}. What is the cluster specific frequency in starbursting irregulars such as 
NGC~1705 and NGC~1569? From the results of  \cite{bill02} and A08 we have in NGC~1705 only 2 clusters (including the SSC) satisfying the \cite{lari99} definition; this implies $T_N\sim0.87$, adopting $M_B(NGC~1705)=-15.9$. 
From Table~2 of \cite{hunter00}, and adopting the new distance of 2.96 Mpc by Grocholski et al. (in preparation), we have  23 
YMCs in NGC~1569, which imply a frequency $T_N\sim1.81$ ($M_B(1569)\sim-17.76$), higher than those of the most rich spirals in the  \cite{lari00} sample.
Hence, NGC~4449 is more active than average in forming young massive clusters, but definitely less than other starburst dwarfs. 

\cite{gel01} introduced a new definition of ``populous clusters'' and ``super star clusters'' to be those with $-10<M_V<-11$ and $M_V<-11$ at 10 Myr, respectively, while \cite{bill02} adopted less stringent limits of $-9.5$ and $-10.5$ in $M_V$ at 10 Myr. These definitions are somewhat model dependent, since the age correction to 10 Myr depends on the particular model used. 
We provide in Table~\ref{cluster_class} the number of populous and super star clusters identified with the \cite{gel01} definition, assuming different metallicities (Padova models). For each metallicity, the range in numbers reflects the uncertainty in age. The $M_V$ at 10 Myr  is given  in Col.~12 of Table~1. 
We identify 4 to 12 young ($<1$ Gyr) SSCs, and a conspicuous population of old ($>$1 Gyr) SSCs (21 to 27) that could be massive globular clusters. The Z$=$0.02 metallicity provides the lowest number of SSCs. 
Assuming a metallicity of Z$=$0.004 or Z$=$0.008, we end up with a minimum number of 7 young SSCs.  
This is larger than what was found by \cite{gel01}, who  identified in NGC~4449 three young SSCs in addition to the central one (their cluster IDs 31, 27 and 47). These three clusters correspond to our clusters  
43, 48, and 18, which we classify as young SSCs as well. Cluster 43 is located just $\sim$ 20 pc east from the central SSC, while cluster 48 is $\sim$ 150 pc west. Cluster 18 is located 
N-E at $\sim$1.3 kpc from the galaxy center, in the region of young SF. 
Our additional four young SSCs are cluster 6, fairly close to cluster 18; cluster 36, which exhibits a high ellipticity and is located  $\sim$ 250 pc east of the central SSC; 
cluster 66 and cluster 67, $\sim$ 900 pc S-W and $\sim$1.1 kpc south from the galaxy center, in proximity of the stream of blue stars extending south. 
Thus all young SSCs are located in regions of active star formation. The old SSCs are instead more homogeneously distributed over the NGC~4449 field of view, following the same distribution of the old stars in Fig.~\ref{starclu}.
Several populous clusters, from 8 to 22, are also identified in our images.

\begin{deluxetable}{llcccccc}
\tabletypesize
\small
\tablecolumns{8}
\tablecaption{Cluster classification}
\tablehead{
\colhead{$Z_{SSP}$} & \colhead{$N_{cl}$\tablenotemark{a}} & \colhead{$N_{-10..-11}$ \tablenotemark{b}} & \colhead{$N_{-10..-11}$ \tablenotemark{b}} &  \colhead{$ND_{-10..-11}$\tablenotemark{c}} 
 & \colhead{$N_{<-11}$ \tablenotemark{d}} & \colhead{$N_{<-11}$ \tablenotemark{d}} 
 &  \colhead{$ND_{<-11}$\tablenotemark{e}} \\
\colhead{} & \colhead{} &  \colhead{($<$1 Gyr)} &  \colhead{($>$1 Gyr)} & \colhead{} & \colhead{($<$1 Gyr)} &  \colhead{($>$1 Gyr)} & \colhead{} \\
\colhead{} & \colhead{} &  \colhead{} &  \colhead{} & \colhead{[$kpc^{-2}$]} & \colhead{} &  \colhead{} & \colhead{[$kpc^{-2}$]} 
} 
\startdata
0.004 &   81 & 18$-$18 &  1$-$4 &  0.7$-$0.8 &  7$-$12 &  24$-$27 & 1.2$-$1.5 \\ 
0.008 &   81 & 13$-$15 &  3$-$6 &  0.6$-$0.8 &  7$-$11 &  21$-$24 & 1.1$-$1.3 \\
0.02  &    81 &   6$-$7  &   2$-$3 &  0.3$-$0.4 &  4$-$5   &  24$-$25 & 1.1$-$1.1  \\ 
\enddata
\tablenotetext{a}{Total number of candidate star clusters, not including the central super star cluster which is saturated in our 
ACS images.}
\tablenotetext{b}{Number of populous star clusters, taken to be those with $-10<M_{V}<-11$ at 10 Myr. Notice 
that the definition is slightly different than that adopted by Billett et al.(2002), Table~2. The range in numbers reflects the uncertainty in age.}
\tablenotetext{c}{Number density of populous clusters found within the $\sim$ 26 kpc$^2$ area covered by our ACS data.} 
\tablenotetext{d}{Number of super-star clusters, taken to be those with $M_{V}$ at 10 Myr brighter than $-11$. Notice 
that the definition is slightly different than that adopted by Billett et al.(2002), Table~2. The range in numbers reflects the uncertainty in age.}
\tablenotetext{e}{Number density of super-star clusters found within the $\sim$ 26 kpc$^2$ area covered by our ACS data.} 
\label{cluster_class}
\end{deluxetable}

\subsection{Correlations between cluster properties. \label{cluster_correlations}}

We investigated the presence of correlations between the cluster properties. 
We find significant correlations only for the red ($V-I>0.9$, $B-V>0.5$) cluster population 
between the effective radius $R_{e}$ and magnitude, $R_{e}$ and  cluster mass, and ellipticity $\epsilon$ and magnitude
(see Fig.~\ref{correla}). The effective radius tends to increase with the cluster magnitude, meaning that brighter clusters are more compact, while 
fainter clusters tend to be more extended. The Spearman correlation coefficient is $r_s=0.63$ with 25 degrees of freedom, implying 
a probability P$\sim$0.0004 that a correlation is not present. 
The correlation between $R_{e}$ and magnitude seems, at least in part, driven by mass: in fact we find that more massive clusters are also more compact. With the masses derived from the Z$=$0.004 Padova models, 
the correlation coefficient is $r_s=-0.54$, implying a probability P$\sim$0.004  that a correlation is not present. Our conclusions do not change if we adopt the masses derived with the GALEV models, or with different metallicities. 
Finally, we find an anti-correlation between magnitude and ellipticity, in the sense that brighter clusters tend to be more elliptical.
The correlation coefficient is $r_s=-0.49$, implying a  probability P$\sim$0.009  that a correlation is not present. 

In an early study, \cite{vdb83} noticed that in the LMC more luminous clusters tend to be more flattened, but the data now available 
show no evidence for such a correlation \citep{vdb08}. This result is true  both for globular clusters and for younger clusters. 
For what concerns the SMC, \cite{vdb08} pointed out that available data are too scant to determine if there is a correlation between the luminosity and the flattening 
of clusters, but noticed that the four brightest SMC clusters are all very flattened having $<\epsilon>=0.26$. 
Interestingly, $\omega$Cen, the brightest globular cluster of the MW, 
is also one of its most elongated ones (see the Harris catalog at {\tt http://www.physics.mcmaster.ca/$\sim$harris/mwgc.dat}).

\section{Cluster disruption and infant mortality \label{cluster_disruption}}

The term ``infant mortality'' was initially introduced by \cite{lada03} to indicate 
that 90\% of the embedded clusters in the solar neighbourhood do not survive the 
gas expulsion phase. According to theory, the effects of gas expulsion should be largely over 
within $\sim$20 Myr \citep{good06}. 
\cite{fall05} and \cite{whit07} introduced the concept of 
``long-term infant mortality'', finding that in the  Antennae 90\% of the clusters, independent of their mass,  
are disrupted every dex of age, a process that continues between a few Myr to $\sim$ 100 Myr, and possibly up to $\sim$1 Gyr. 
The same fraction (90\%) of infant mortality was claimed to be present also in other 
galaxies, such as the SMC \citep{chandar06}, the LMC \citep{chandar10a}, and M83 \citep{chandar10b}.
\cite{pellerin07} claimed  significant infant mortality in the spiral NGC~1313 from the fact that O stars were more ``clustered'' 
together, over $\sim$100 pc, than B stars (notice however that these scales imply that this is unlikely to be caused by traditional ``infant mortality'').
These results are in contradiction with other studies, which instead do not find evidence for long-term infant mortality \citep{gb08,bastian09}.

The most common way to study cluster disruption is to use the cluster age
distribution, which, however, can be heavily affected by incompleteness. 
\cite{gb08} proposed a new method of studying 
cluster disruption based on the relation between the most 
massive cluster, M$_{max}$, and the age range sampled.
\cite{gb08} showed that, assuming that clusters are stochastically sampled from a power-law cluster 
initial mass function with index $-2$ \citep{zf99, bik03b,degri03}, 
and that the cluster formation rate (CFR) is constant, 
$M_{max}$ scales with the age range sampled, 
such that the slope in a $\log(M_{max}$) vs. $\log(age)$ plot 
is equal to unity. In the scenario of mass independent (MID) cluster disruption, 
the slope turns out to be $[1 + log(1-f_{MID})]$, where $f_{MID}$ is the fraction of clusters destroyed in each age dex.
Thus the slope becomes flatter than 1 as cluster disruption is more important.
 
We investigated the $\log(M_{max}$) vs. $\log(age)$ relation for the young (age$<$1 Gyr) clusters 
in NGC~4449. Our results are shown in Fig.~\ref{mmax_tot}. 
We adopted an age bin of 1 dex starting from log(age[yr])$=$6. This is the same bin adopted 
by \cite{chandar10b}, and is larger than the 0.5 dex bin adopted by \cite{gb08}.
Given the bin size, we will not be able to characterize the 
``short term infant mortality'' within the first $\sim$10 Myr. 
The slope of the $\log(M_{max}$) vs. $\log(age)$ relation was computed for the three metallicities 
Z$=$0.004, 0.008 and 0.02, with the GALEV and with the Padova models. 
We assumed both no internal reddening 
(top panels), and an $E(B-V)$ as high as 0.2 (middle panels). 
Notice that this value is twice the 
average $E(B-V)\sim0.1$ derived by \cite{hg97} in NGC~4449 HII regions, and 
should thus be considered an upper limit. 
The slopes exhibit some dependence on metallicity 
and on the adopted models (Padova or GALEV). If we average over the 
 three metallicities and over the two models, we obtain $0.76\pm0.07$ in the case 
of no internal reddening, and $0.58\pm0.12$ in the case of $E(B-V)=0.2$. 
These slopes imply cluster disruption fractions of  $f_{MID}=0.42_{-0.10}^{+0.09}$ and 
$f_{MID}=0.62_{-0.12}^{+0.09}$, significantly lower than the 90\% 
long term infant mortality found by other authors.

Our conclusions rest on the assumption that the CFR is constant. 
However, what if this is not the case? We can assume that the CFR follows 
the SFR in the field. From Fig.~9 and~10 of \cite{mcq10} we derive SFRs of $\sim$0.98, 
$\sim$0.36, and $\sim$0.19 M$_{\odot}$/yr in the  $<$10 Myr, (10$-$100) Myr, and (100$-$1000) Myr age bins,
meaning that NGC~4449 has been more active at recent epochs than in the past.
We thus ``correct'' the  $M_{max}$ values for the fact that the SFR was not constant over the past 1 Gyr. 
Since $M_{max} \propto N_{bin}$, where $N_{bin}$ is the number of clusters in each age bin 
(see Eq.(7) of \cite{gb08}), and $N_{bin}\propto$ CFR $\propto$ SFR , we re-compute the slopes using  
the normalized $log(M_{max}/C_{SFR})$, where  $C_{SFR}$ is proportional to the SFR in the corresponding age bin  
(bottom panels of Fig.~\ref{mmax_tot}). In this case the average slope is as steep as $1.12\pm0.07$,  
as a consequence of the more intense average SF experienced by NGC~4449 over the last $\sim$100 Myr,
and in particular over the last $\sim$10 Myr. This is consistent with no cluster disruption.

Finally, we should comment on the fact that our catalog does not include the central SSC. 
\cite{gel01} derived for the SSC an age between 8$-$15 Myr 
and a mass of $2\times10^5 M_{\odot}$, while \cite{bok01} derived an age between 6$-$10 Myr and a lower limit 
for the mass of $4\times10^5 M_{\odot}$.  Thus, if included in our $M_{max}$ analysis, the SSC would provide significantly flatter slopes.
However, \cite{gel01} showed the the central SSC in NGC~4449 is a very peculiar object: it consists of a 
bright condensation 0.1'' in diameter sitting in the middle of an elongated, bar-like structure.
The minor-to-major axis ratio of the structure is 0.6, while the cluster is essentially round in the outer 
parts. \cite{gel01}  suggested that this structure is the result of the interaction scenario 
that also produced the counterrotating gas systems and extended gas filaments and streamers  
in NGC~4449. In particular, the peculiar structure at the center of the SSC could be the 
debris from cannibalism of a smaller companion that has fallen into the center of the galaxy.
In this context, the central SSC in NGC~4449 appears a peculiar object whose formation is not 
comparable to the formation of the population of ``normal'' clusters. 
More generally, it has been well documented that nuclear star clusters in galaxies (``nuclei'') 
tend to have distinct properties from globular clusters in general \citep[e.g.,][ and references therein]{vdm07}. 
This is not unexpected, given the special location at the center of a gravitational potential well. This further motivates our decision to exclude the central SSC in NGC~4449 from our analysis of the general cluster population.

\section{Results and discussion \label{discussion}}

We have presented a study of 81 candidate star clusters in NGC~4449, 37 of which 
in common with a previous study by \cite{gel01}, based on our ACS/WFC 
F435W ($\sim$B), F555W ($\sim$V), F814W ($\sim$I), and F658N ($H\alpha$) images.
Our ACS images are saturated in the center of the galaxy, preventing an analysis  of the central super star cluster (SSC), already studied by others authors \citep{gel01,bok01}. 
The spatial distribution of the clusters appears to follow that of the resolved stars in NGC~4449.  

We derived the cluster intrinsic properties such as the intrinsic effective radius $R_{e}$ and the ellipticity $\epsilon$, as well as the integrated 
magnitudes. $R_{e}$ is found to vary from $\sim$ 1 pc to $\sim$ 9 pc, with a median value of $\sim$ 3.2 pc, while the ellipticity ranges from 
$\epsilon \sim 0$ to 0.4, with a median value of $\sim$ 0.1, and with the most elliptical clusters ($\epsilon>0.3$) being all blue ($B-V<0.5$ or $V-I<0.9$). Artificial cluster experiments show that the errors on the size measurements increase from bright to faint and from compact to more extended clusters. 
The error in size is $\sim$1\% for a cluster with $R_{e}\sim$4 pc and V$=$20, and it is as high as $\sim$ 6\% if $V=22.5$. 
 These should be considered lower limits, since they were derived fitting the artificial cluster profiles with the same law assumed to create them.
Very extended clusters ($R_{e}\sim$ 8 pc) suffer strong incompleteness at relatively bright magnitudes, and are easily lost in the most crowded regions.

In a $B-V$ vs. $V-I$ color-color diagram, the candidate star clusters define four different  
populations, a red one with $B-V>0.5$ and $V-I>0.9$ (27 clusters), a blue one with 
$0<B-V<0.5$ and $0<V-I<0.9$ (44 clusters), a very blue one with $B-V<0$, $V-I<0$ (4 clusters), 
and a population of V-bright clusters with $V-I<0$ and $B-V>0$ (6 clusters). 
We find nebular emission associated with 11 clusters out of 81, and specifically with all the 6 V-bright clusters, with 2 of the very blue clusters, and with 3 of the blue clusters. 
SSP models of different metallicities are strongly degenerate in the color-color diagram, thus our results are strongly affected by the age-metallicity degeneracy. However, independent of metallicity, red clusters have ages $\ga$ 1 Gyr, blue cluster $<$ 1 Gyr, and very blue clusters are as young as a few Myr. Likely, the V-bright objects are young (age$\lesssim$ 5 Myr) massive clusters embedded in the ionized gas, as suggested by the fact that the H$\alpha$ emission is coincident with the cluster itself, 
even if we can not exclude that some of them are background emission-line galaxies.
From the B, V, I integrated colors, we are able to derive an age for 8 clusters out of the 11 associated with nebular 
emission. The majority of them have ages of $\sim$ 5-7 Myr, two have ages of $\sim$ 100 Myr, and one is as old as 
$\sim$300 Myr. For the youngest clusters, the H$\alpha$ emission is closely associated with the cluster itself and has the form of an envelope surrounding it, while in the oldest ones the nebular emission is more extended and/or distant from the cluster, likely because it was pushed out by massive star winds and supernovae. These findings are in agreement with the results of \cite{whit11} pointing toward a close connection between the H$\alpha$ morphology/extent and the cluster age.

Notice that clusters as young as a few Myr are numerous in NGC~1569 \citep{hunter00,origlia01}, 
and are possibly present in IZw~18 \citep{contreras11}, but are not found in NGC~1705 \citep{bill02,ann09}. 
If we assume a  Z$=$0.004 metallicity (which is consistent with the HII region abundances and with 
the RGB colors of the resolved stars in NGC~4449), we find that the cluster ages are distributed quite continuously over a Hubble time, and the clusters have masses in the range 
$\sim10^3- 2\times10^6 M_{\odot}$. With the the Z$=$0.02 models we obtain instead a completely different scenario, i.e. a bimodal age distribution strongly peaked at $\sim$ 6 Myr and $\sim$ 1.5 Gyr. This metallicity is a factor $\sim$ 4 higher than the metallicity derived in the 
HII regions, and thus difficult to accept. However, also \cite{gel01} noticed that the ensemble of clusters in NGC~4449 are 
fitted better with cluster evolutionary tracks of somewhat higher metallicity than one would have predicted 
from the nebular oxygen abundance.  
If we consider the results obtained with the  Z$=$0.004 SSP models, we 
find that, within the uncertainties, the cluster age distribution is consistent with the SFH of NGC~4449 
derived by McQuinn et al (2010) from the analysis of the CMDs of the galaxy resolved stars.
Furthermore, young clusters are preferentially located in regions of young star formation, 
while old clusters are distributed over the whole NGC 4449 field of view, like the old stars (although we notice that 
some old clusters follow linear structures, possibly a reflection of past satellite accretion). 
    
These results may sound obvious, if one believes that all stars
originally form within clusters, but actually this is not always the case. For
instance, the well known bimodal SFH of the LMC clusters \citep{pagel98} is completely at odds with that of representative LMC regions 
\citep[see e.g.][and Fig.~7 in Tosi et al.~2006]{smec02}. Thus, also the scenario of a bimodal cluster age distribution can not be excluded a priori. 
We derive a lower limit of  $\sim10^7 M_{\odot}$ for the total mass in clusters. For comparison, the total mass from the SFH by \cite{mcq10} in the same field of view is $3\pm0.5 \times 10^9 M_{\odot}$.
 
 We investigated possible correlations between the cluster properties (magnitude, size, ellipticity, mass and age).
 Significant correlations are found only for the old ($V-I>0.9$, $B-V>0.5$) cluster population 
between the cluster effective radius and the magnitude, the effective radius and the mass, and between the ellipticity $\epsilon$ and the magnitude. Brighter and more massive clusters tend to be more compact, and brighter clusters tend to be also more elliptical. 
\cite{vdb08} noticed that, while in the LMC there is no evidence for a correlation between ellipticity and luminosity, 
the four brightest SMC clusters are all very flattened having $<\epsilon>=0.26$. 
Interestingly, also $\omega$Cen, the brightest globular cluster of the MW, is one of its most elongated ones.

Among the old clusters in our sample, one of the most massive ($M\sim1.7\times10^6 M_{\odot}$) and elliptical ($\epsilon\sim0.2$) one (namely, cluster 77), appears surrounded 
by a symmetric structure of blue stars, which could be tidal tails associated with a dwarf galaxy currently disrupted by NGC~4449. This cluster is treated in more details in an accompanying paper (Annibali et al. in preparation).

Because of the combined effects of incompleteness, cluster mass-function sampling, and possibly mass-dependent cluster mortality
(more massive clusters are more difficult to destroy), clusters are not distributed uniformly in a log mass versus log age plane, but a trend is observed in the sense that the cluster mass seems to increase with the age. 
Following the approach proposed by \cite{gb08}, we used the upper envelope of this distribution, i.e. the maximum cluster mass in different age bins, to test the hypothesis of mass-independent cluster disruption. We do not find evidence for the high (90\%) long-term infant morality 
found by other authors \citep{whit07,chandar06,chandar10a,chandar10b}. If we assume that the cluster formation history follows that of the field stars 
our data are consistent with no cluster disruption within $\sim$1 Gyr, in agreement with  \cite{gb08} and \cite{bastian09}.

As mentioned in the introduction, NGC~4449 is a very active star
forming galaxy, the one with highest star formation rate in the sample of 18
starburst dwarfs recently examined by \cite{mcq10}. This is likely the
cause of its large number of detectable YMCs, confirming the suggestion by
\cite{lari00} of a direct correlation between the number and
luminosity of YMCs and the SF activity of the parent galaxy. We find that  the 
YMC specific frequency in NGC~4449 is higher than the average frequency derived by 
\cite{lari00} in nearby spirals, and also higher than in the LMC, but lower than the frequency 
derived in other starburst dwarfs such as NGC~1705 and NGC~1569.
These results are confirmed by NGC~4449 young cluster LF ($dN(L_V)\propto L_V^{-1.5}dL$) 
flatter than those derived in late-type spirals, and more similar 
to those of blue compact galaxies, perhaps because the environment in these starburst galaxies has 
favored the formation of massive clusters.
Furthermore, NGC~4449 seems to host a significant number of SSCs: adopting the definition of SSC to be a cluster 
with M$_V<-11$ at 10 Myr, introduced by \cite{gel01}, we find that there are 7 young (age$<$1 Gyr) SSCs in NGC~4449, 
besides the central one. Three of these clusters were already identified by  \cite{gel01} from WFPC2 data. 
All the young SSCs are located near the galaxy center or in regions of active SF.

Of particular interest is the comparison of the cluster properties with
the SFR density, which is a useful physical quantity to rank the SF activity of
galaxies of different mass and size. 
From \cite{mcq10}, the average SFR over the last $\sim$100 Myr is $\sim$0.42 $M_{\odot}/yr$, and from Fig.~\ref{starclu} of this paper we estimate that stars younger than $\sim$100 Myr are distributed over an area of $\sim$7.85 kpc$^2$. This provides a 
SFR density of  $\sim$0.05  $M_{\odot}/yr$/kpc$^2$. On the other hand, we count 15 clusters (including the 
central SSC) with ages $\lesssim$100 Myr, which imply a cluster density of $\sim$1.9/kpc$^2$  over the same area.

For NGC~1705, \cite{ann09} give 9 clusters (including the central SSC) younger than $\sim$100 Myr 
concentrated in the central $26^{\prime\prime} \times 29^{\prime\prime}$ area, corresponding, at a distance of 
5.1 Mpc \citep{tosi01}, to $\sim$642$\times$717~pc$^2$. This provides a cluster density of 19/kpc$^2$. 
To compute the SFR density, we sum up the rates in Regions~7, 6 and 5 in Table~3 of \cite{ann09} and 
divide the result by the total area, which gives $\sim$0.08 $M_{\odot}/yr$/kpc$^2$.
Thus, despite the fact that the SFR density in NGC~1705 is a factor $\sim$2 higher than that in NGC~4449, the cluster specific frequency is 
$\sim$ a factor 10 higher.

NGC~1569 is the dwarf galaxy with the highest SFR density: from HST/NIC2 data, \cite{ange05} derived an 
average SFR of $\sim2.5~M_{\odot}yr^{-1}kpc^{-2}$ over the past 100 Myr, assuming a distance of $\sim$2.9 Mpc.
NGC~1569 is also known to host 4 SSCs, and several normal clusters  \citep{hunter00,origlia01}. 
\cite{hunter00} identified 48 clusters in their WFPC2 images, 28 of which fall within the $19^{\prime\prime} \times 19^{\prime\prime}$  
NIC2 field of view. From the cluster photometry in their Table~2,  and using the Padova models, we estimate that 
25 of them have ages $\lesssim$100 Myr. This implies a cluster density of $\sim$357/kpc$^2$, and a cluster/SFR density ratio of $\sim$142,
a factor $\sim$4 higher than in NGC~4449.

These results agree with the findings of \cite{lari00}, according to which NGC~1705 and 
NGC~1569 deviate from the cluster specific luminosity $T_L(U)$--SFR density $\Sigma_{SFR}$ relation defined by star forming galaxies,  exhibiting very large  $T_L(U)$ for their  $\Sigma_{SFR}$. 

Another interesting correlation is the one found by  \cite{lar02} between the brightest cluster V-band absolute magnitude, $M_V^{brightest}$, and the galaxy 
SFR,  suggested to be caused by size-of-sample effects. This trend has been used to argue for a universality of cluster formation, i.e. stochastic sampling from a universal mass function. \cite{bastian08} showed that the absolute magnitude of the brightest young cluster in a galaxy is a good tracer of its current SFR; he also showed that the fact that dwarf/irregular starburst galaxies often lie signiÞcantly above the relation can be explained in terms of their SFH proceeding in short bursts. 
From our Table~1, the brightest cluster in NGC~4449 has  $M_V=-10.75\pm0.15$, which falls significantly ($>5\sigma$) below the relation 
$M_V^{brightest}=1.87(\pm0.06) \times \log SFR -12.14(\pm0.07)$ \citep{weidner04}, adopting a current SFR of  $\sim 1 M_{\odot} yr^{-1}$ \citep{mcq10}.
On the other hand, the central SSC, with a magnitude  of $M_V=-12.54\pm0.15$ (adopting the value from \cite{gel01} and correcting for  a distance modulus of $DM=27.91\pm0.15$) turns out consistent with the relation within $3\sigma$.

\acknowledgments

F. A. would like to thank P. Anders and R. Kotulla for clarifications about the GALEV models, 
and A. Bressan and P. Pessev for useful discussions.  We thank A. Grocholski for providing 
the results about NGC~1569 before publication.
We thank the anonymous referee for his/her useful comments/suggestions which helped to 
improve the paper.

Support for proposal \#10585 was provided by NASA
through a grant from STScI which is operated by
AURA    Inc.    under NASA contract NAS 5-26555.
FA and MT have received partial financial support from ASI, through contract COFIS
ASI-INAF I/016/07/0 and contract ASI-INAF I/009/10/0.





\appendix

\clearpage

\begin{figure}
\epsscale{1.}
\plotone{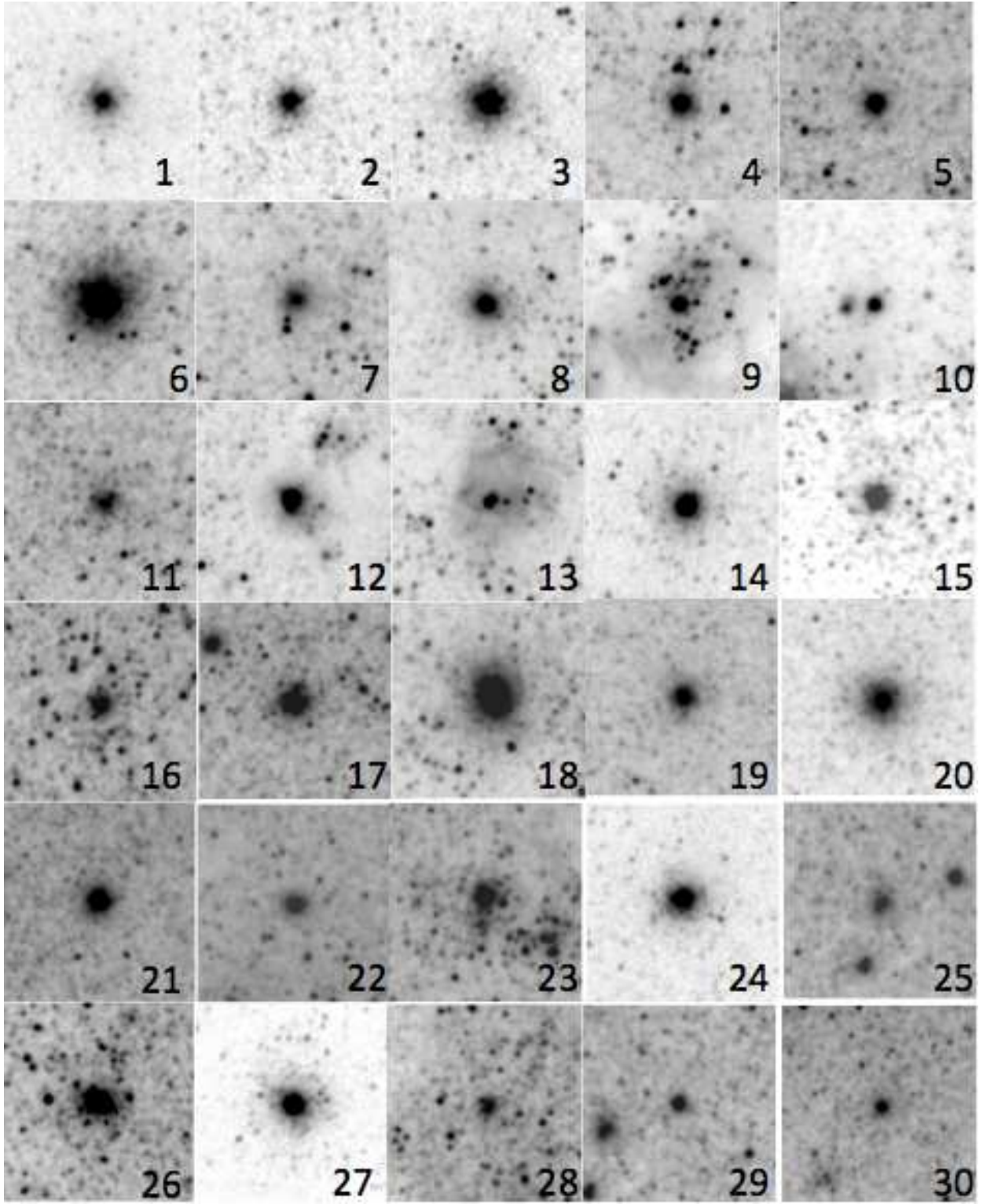}
\caption{(4.5$\times$4.5) arcsec$^2$ F555W images of the cluster candidates.
The big cluster right of cluster number 43 is the central SSC, which is saturated in our images.
 \label{clustera}}
\label{image}
\end{figure}

\addtocounter{figure}{-1} 
\begin{figure}
\epsscale{1}
\plotone{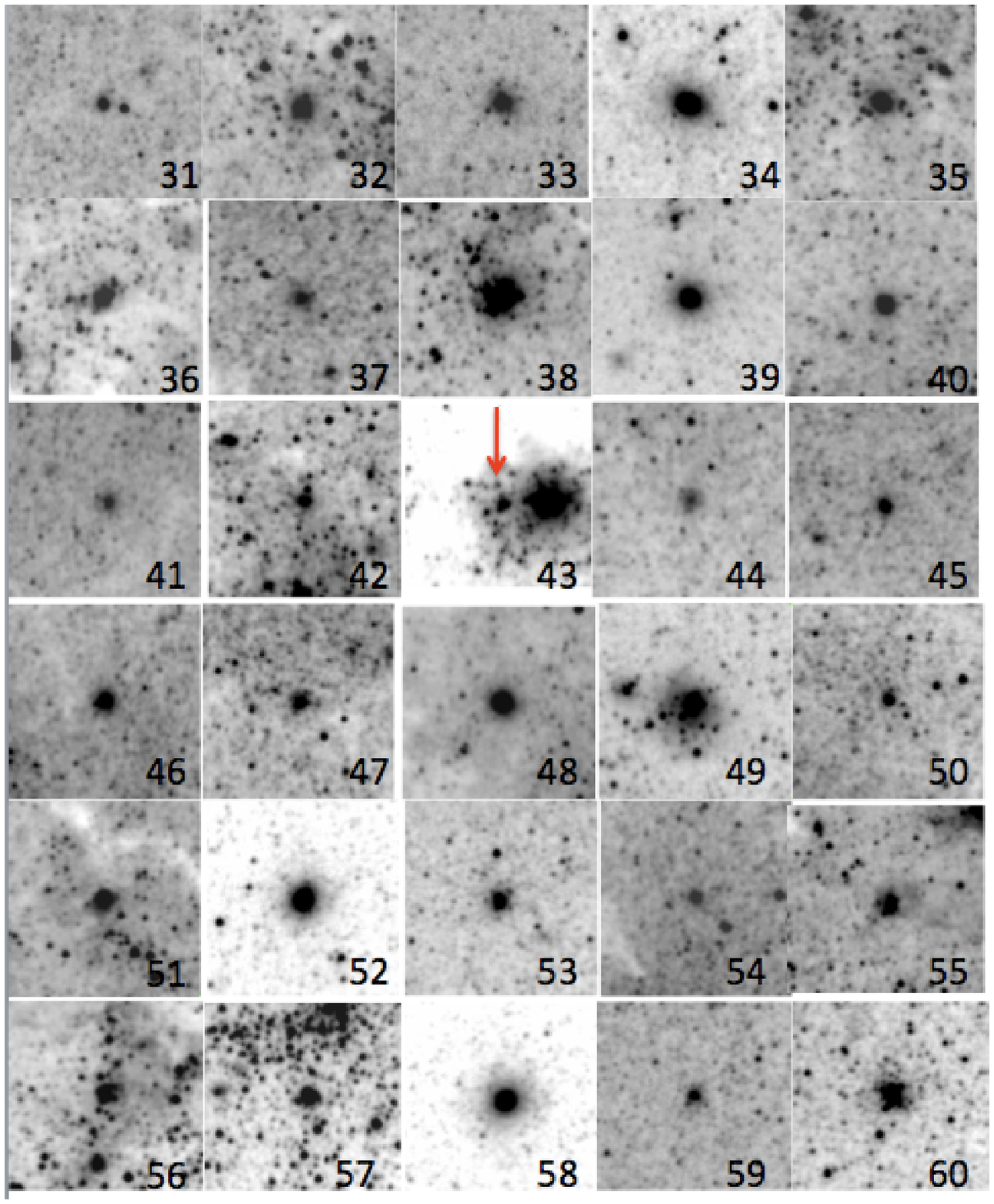}
\caption{continued }
\end{figure}

\addtocounter{figure}{-1} 
\begin{figure}
\epsscale{1}
\plotone{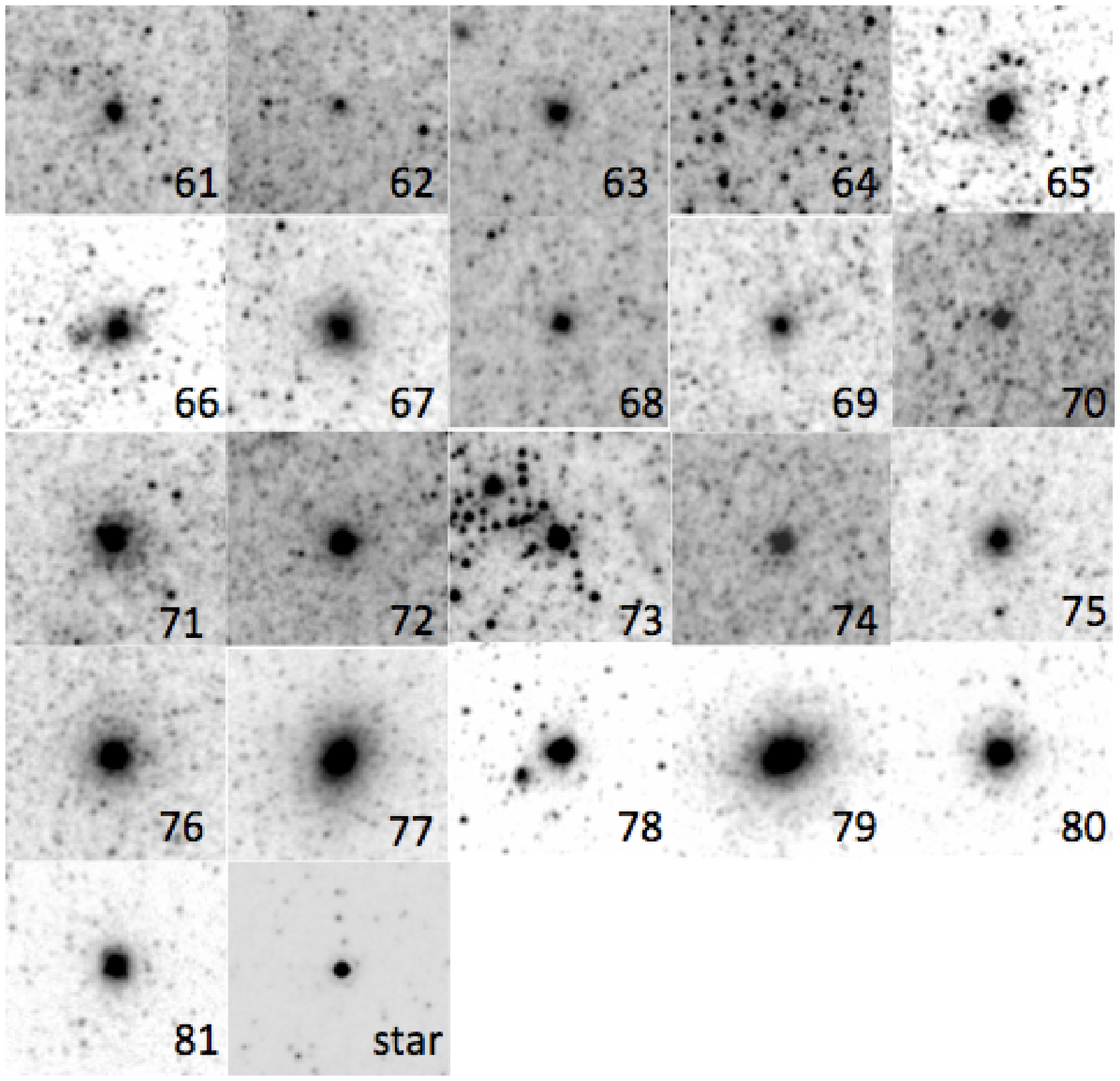}
\caption{continued }
\end{figure}

\begin{center}
\begin{figure}
\epsscale{0.9}
\plotone{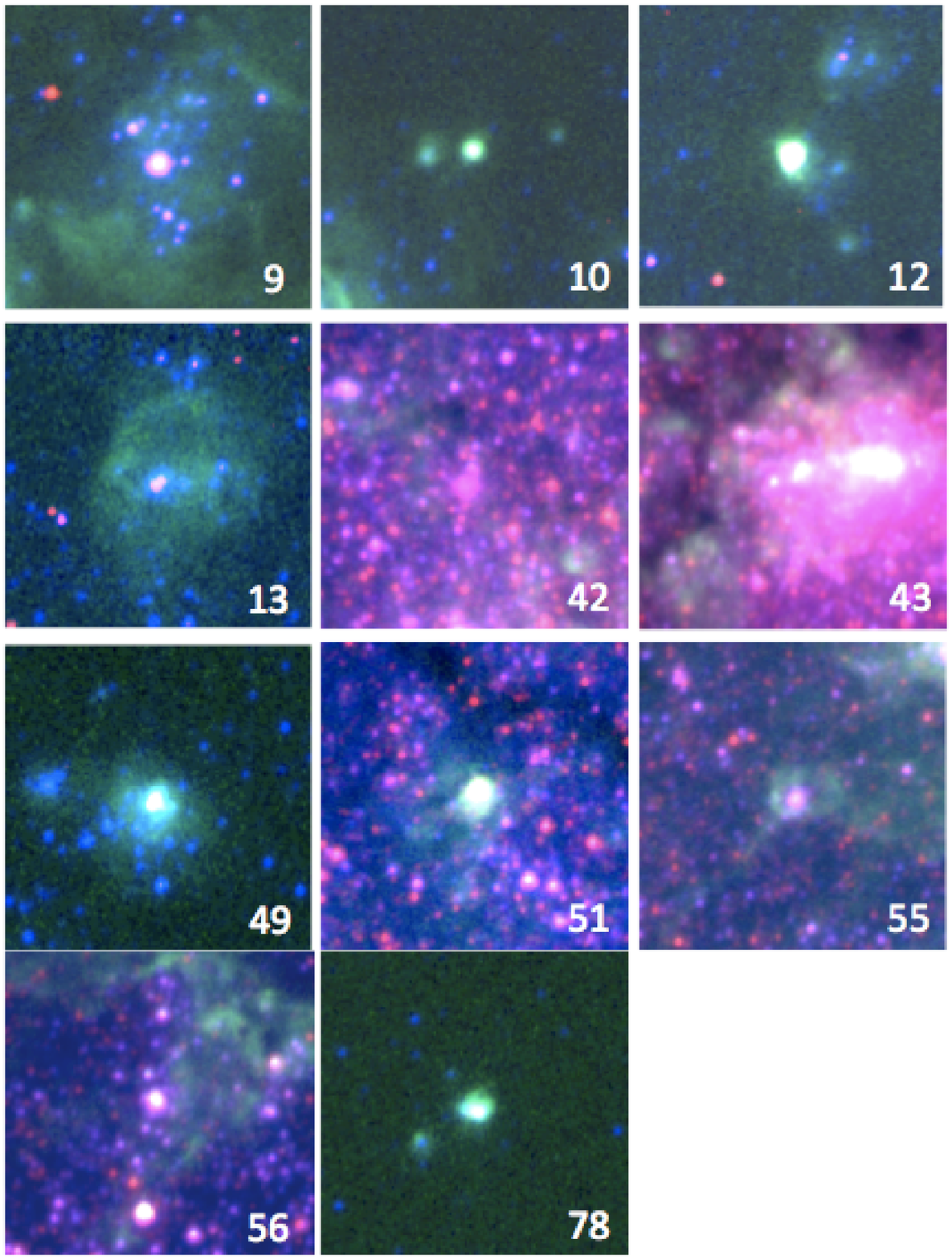}
\caption{Color-composite images (F435W$=$blue, F658N$=$green, F814W$=$red) of the clusters 
with associated H$\alpha$ emission. For each cluster, the field of view is  (4.5$\times$4.5) arcsec$^2$. 
\label{clu_ha}}
\end{figure}
\end{center}

\begin{center}
\begin{figure}
\epsscale{1.}
\plotone{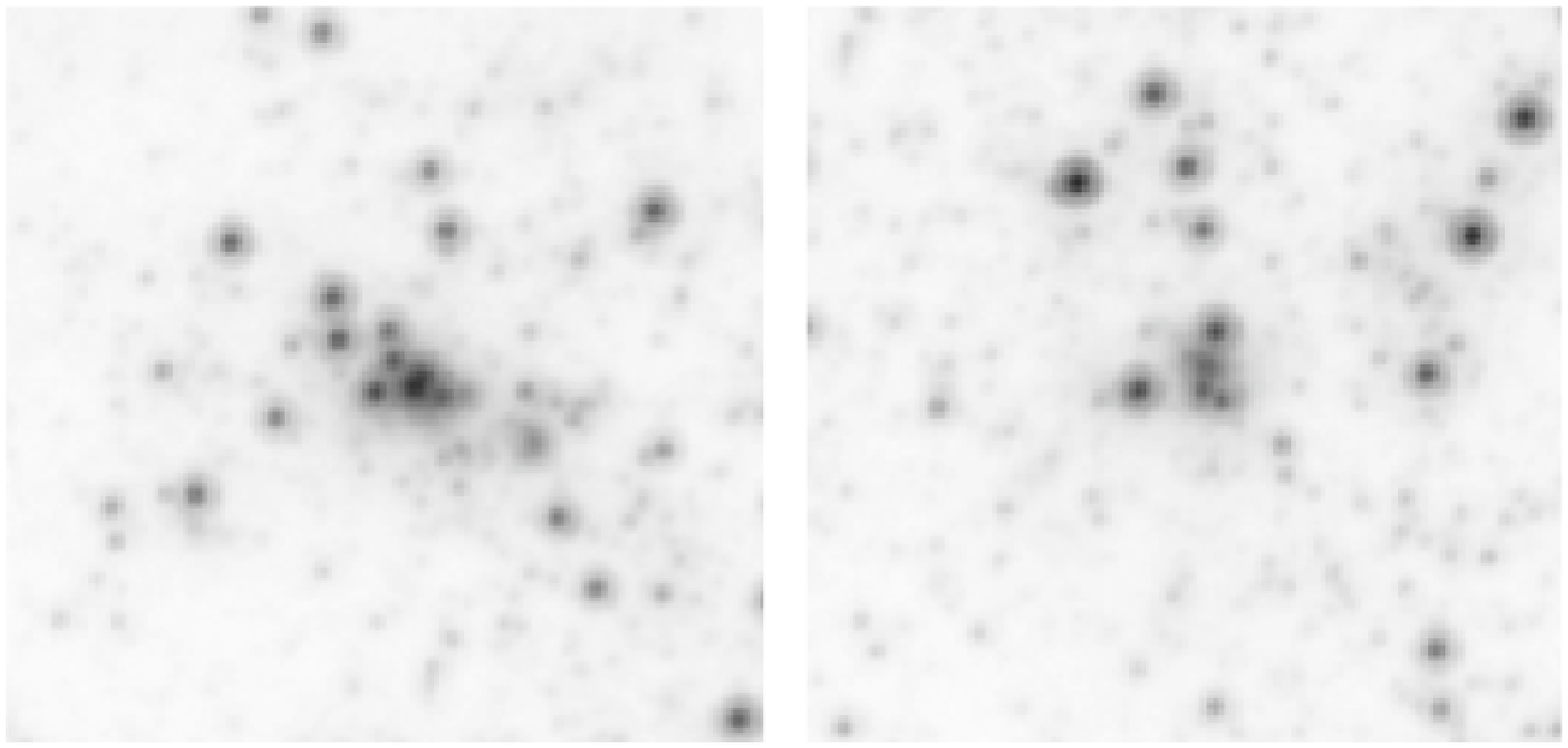}
\caption{(4.5$\times$4.5) arcsec$^2$ F814W images of two clusters classified by \cite{gel01} but considered by us to be ``stellar associations'' (see Section~\ref{selection}).
\label{2exgel}}
\end{figure}
\end{center}

\clearpage
\begin{figure}
\epsscale{1}
\plotone{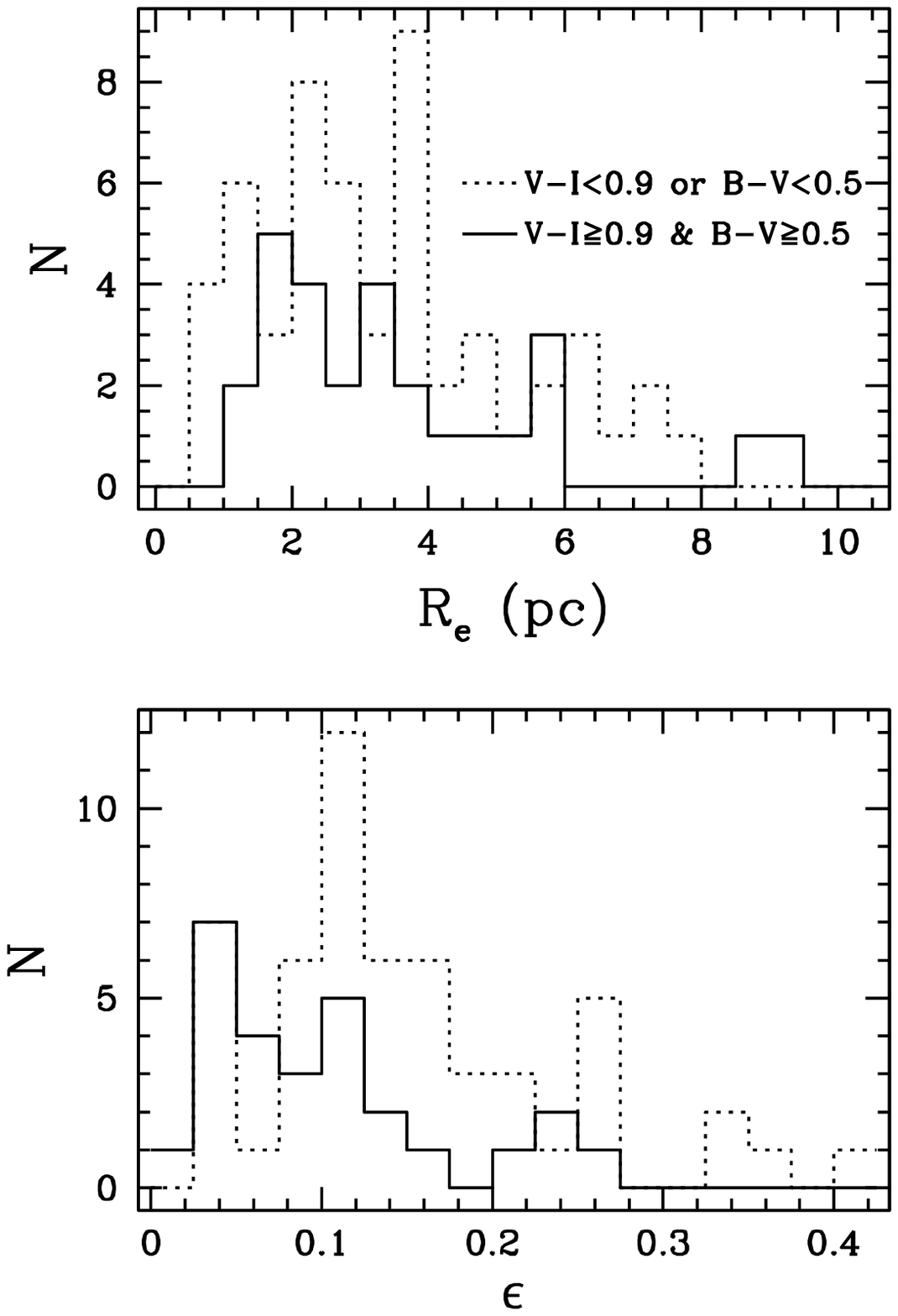}
\caption{Distributions of the intrinsic effective radii $R_e$ (top panel) and of the ellipticities $\epsilon$ 
(bottom panel) for blue 
($m_{F555W}-m_{F814W}<0.9$ or $m_{F435W}-m_{F555W}<0.5$) and red ($m_{F555W}-m_{F814W}\ge0.9$ and $m_{F435W}-m_{F555W}\ge0.5$) clusters. \label{reffeps}
}
\end{figure}

\begin{figure}
\epsscale{1}
\plotone{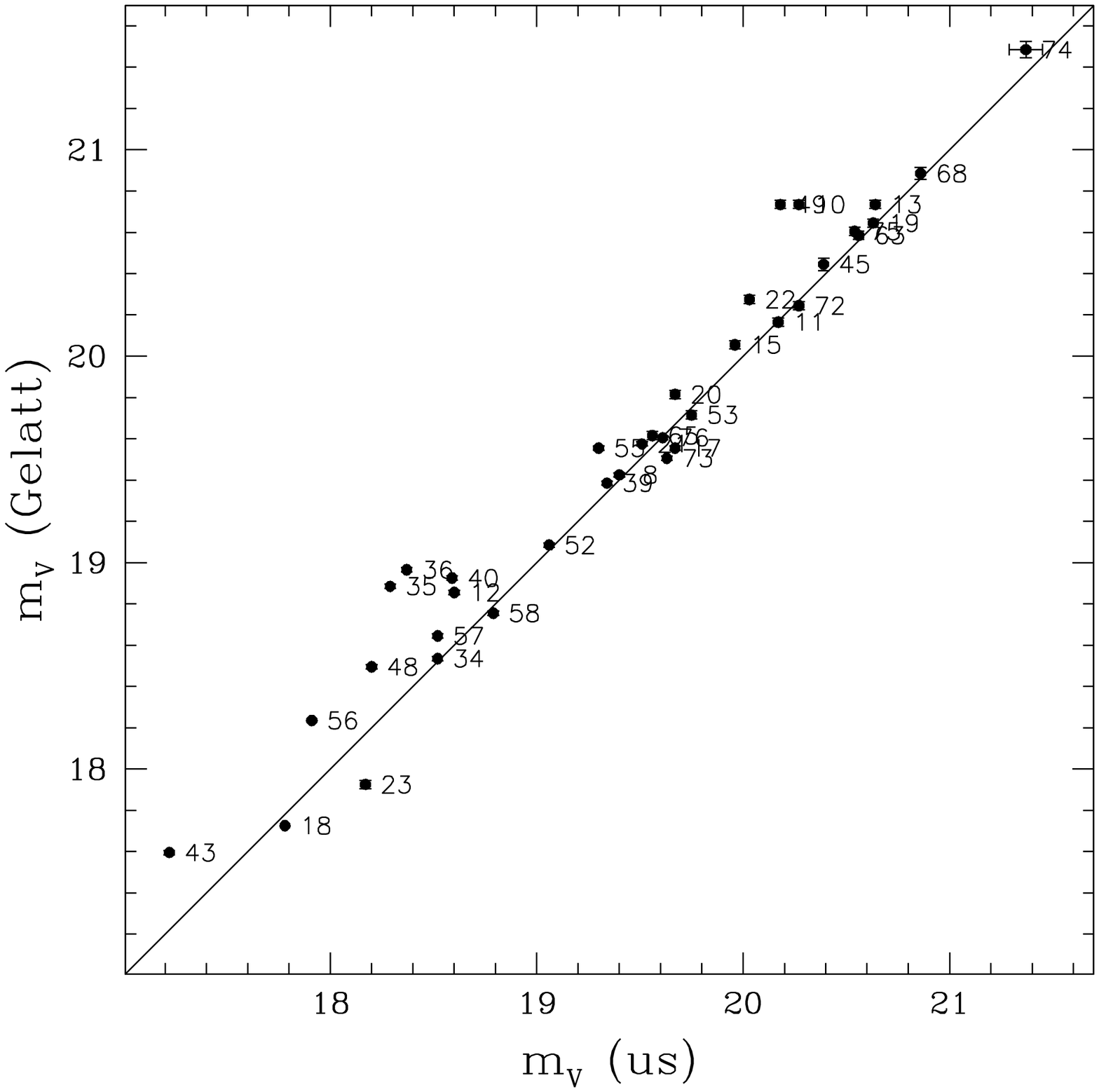}
\caption{Comparison of the observed Johnson-Cousins V magnitudes measured by us and by \cite{gel01} for the 
clusters in common. The solid line is the one-to-one relation. \label{gelattus}
}
\end{figure}

\begin{figure}
\epsscale{1}
\plotone{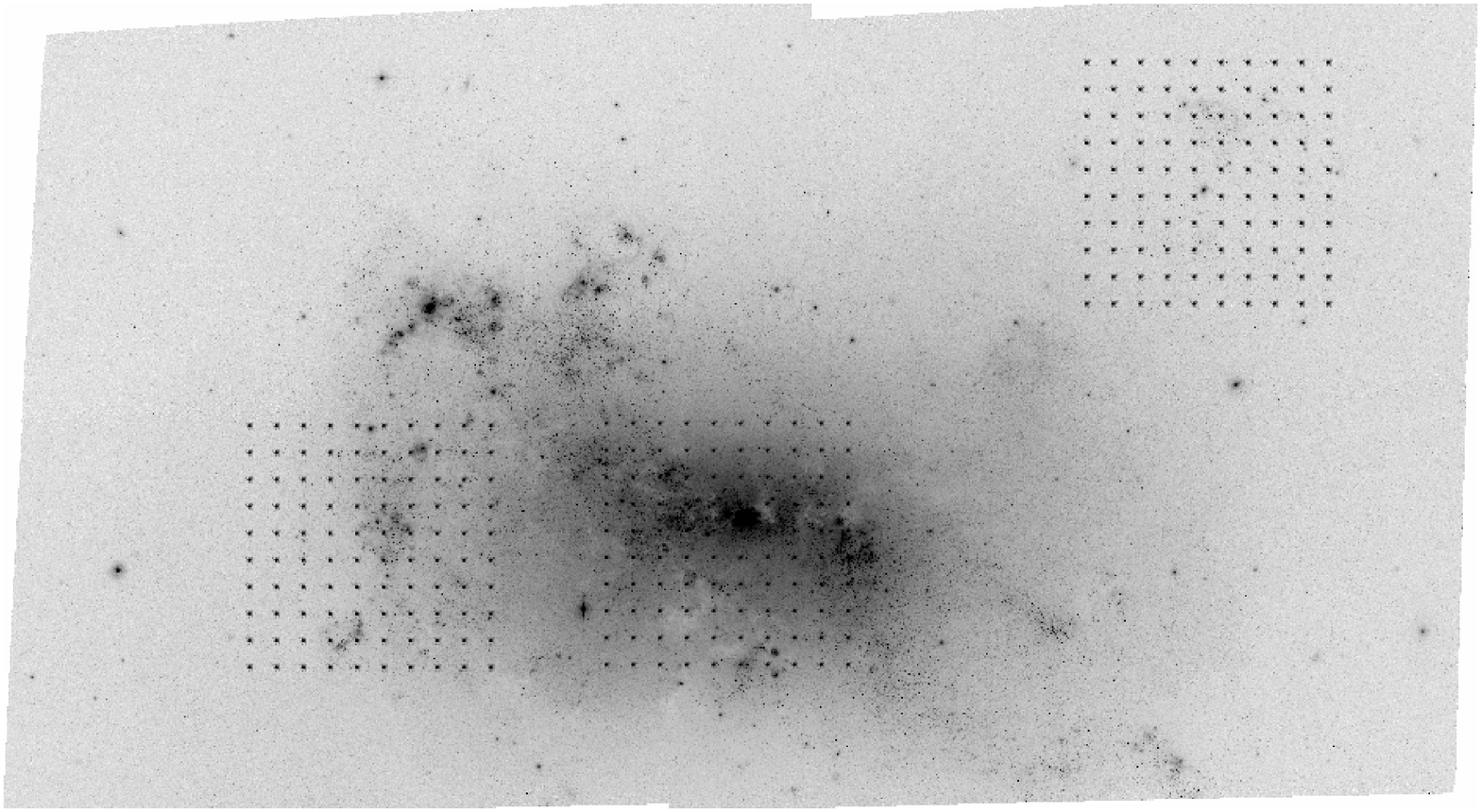}
\caption{Artificial clusters added to the (380 $\times$ 200) arcsec$^2$ mosaicked ACS  F555W image in 
three fields of different crowding. From left to right are Field 2 (medium crowding), Field 1 (high crowding), 
and Field 3 (low crowding). \label{image_compl}
}
\end{figure}


\begin{figure}
\epsscale{1}
\plotone{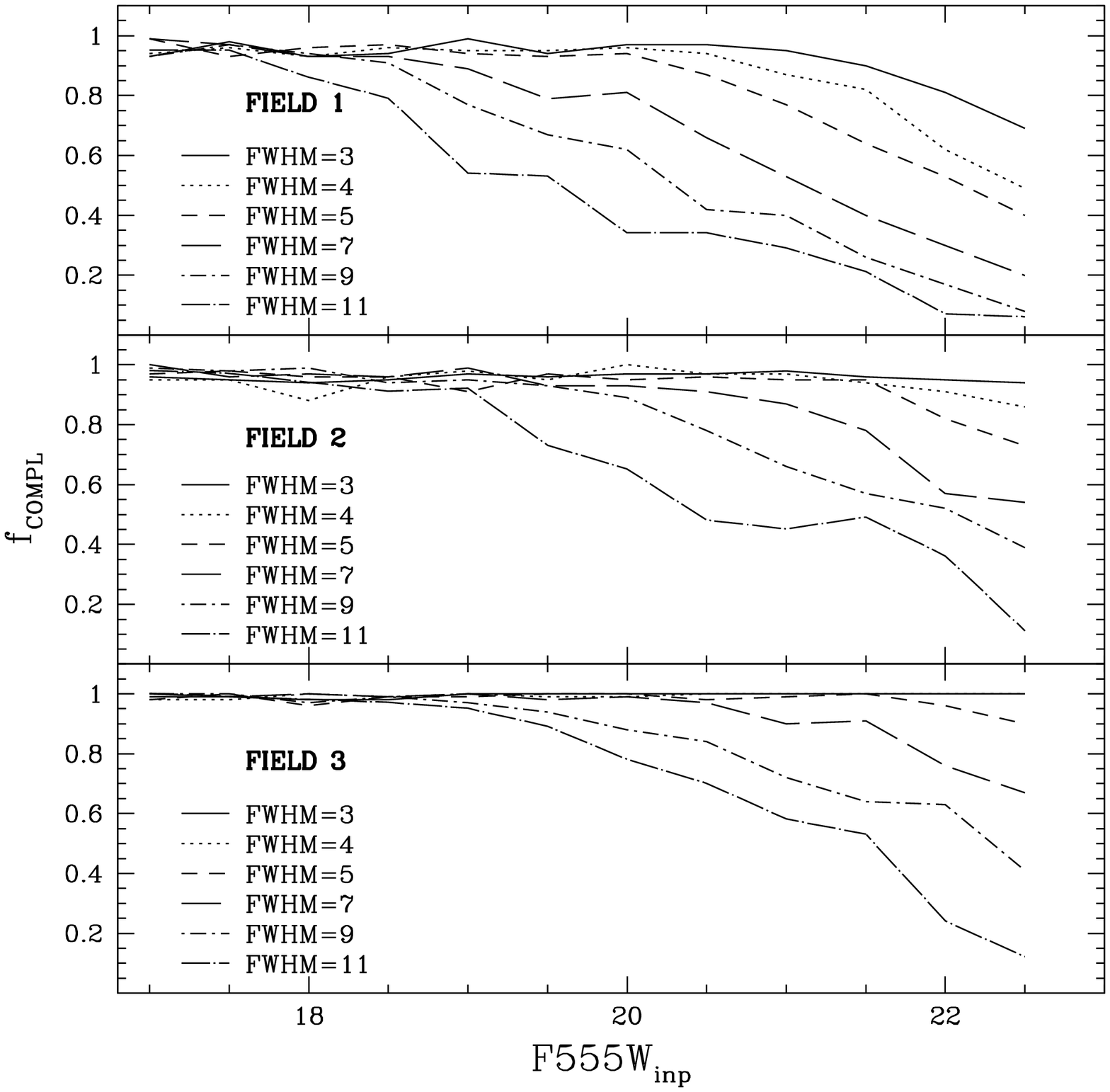}
\caption{Completeness in the F555W  band  for artificial clusters with
a {\it MOFFAT15} intrinsic profile and FWHM of 3, 4, 5, 7, 9 and 11 pixel (1 pixel=0.035'') 
in the three test fields (the crowding increases from Field 3 to Field 1). \label{compl}
 }
\end{figure}


\begin{figure}
\epsscale{1}
\plotone{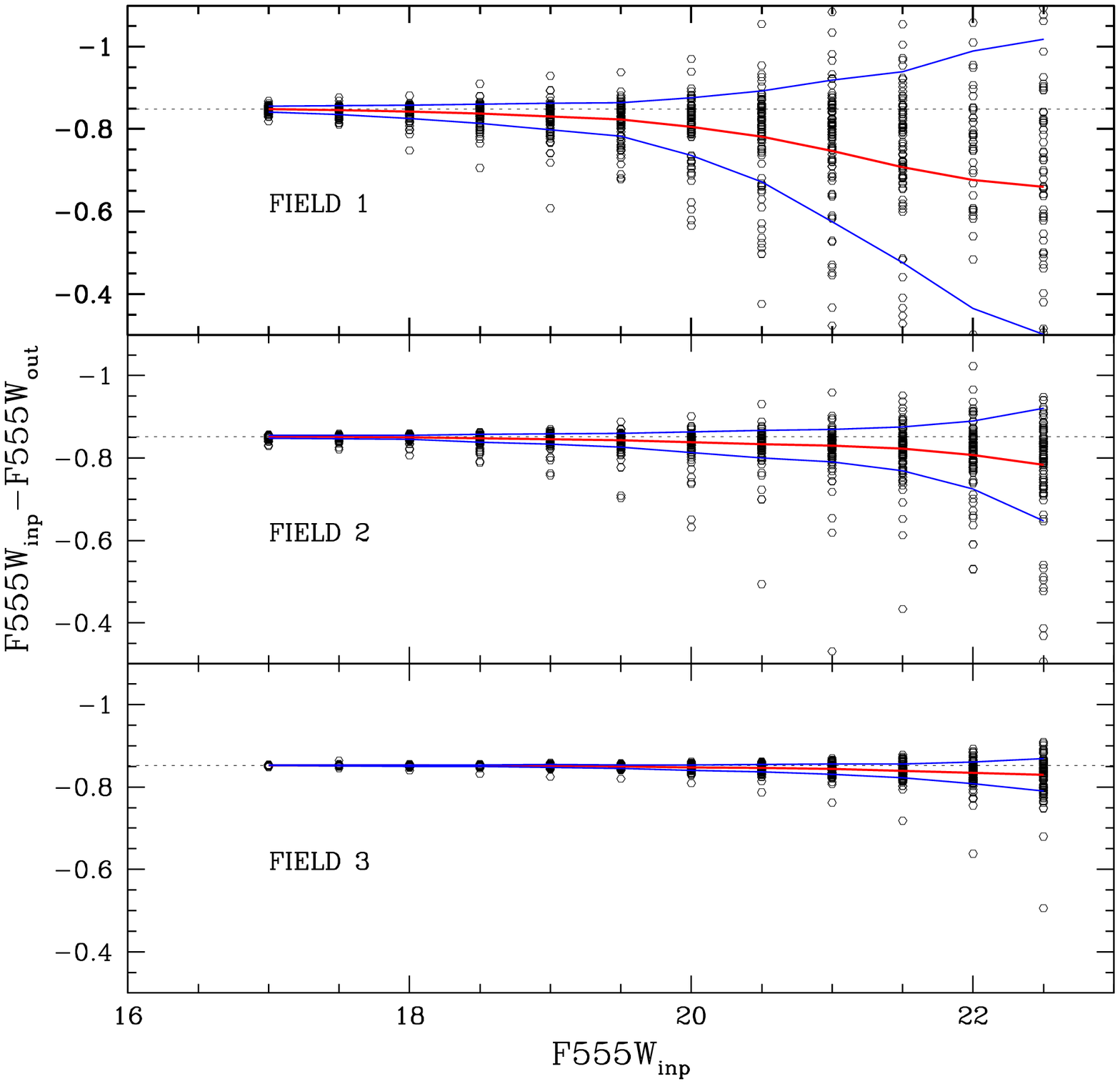}
\caption{Example of input minus output magnitudes in F555W 
for artificial clusters with an intrinsic {\it MOFFAT15} profile and an 
intrinsic FWHM$=$3 pixels ($\sim$0.1'').  The output magnitudes were derived 
through photometry within a $R_{phot}$ aperture (see Section~\ref{photometry} for 
details), while $F555W_{inp}$ is the total cluster magnitude.
The average mag(inp)$-$mag(out) value at different magnitudes 
(central red line) is the aperture correction from  $R_{phot}$ to ``infinite'' to be applied to the photometry. 
The $\pm 1 \sigma$ levels (blue external lines) quantify the photometric errors. \label{errors}
 }
\end{figure}

\begin{figure}
\epsscale{1}
\plotone{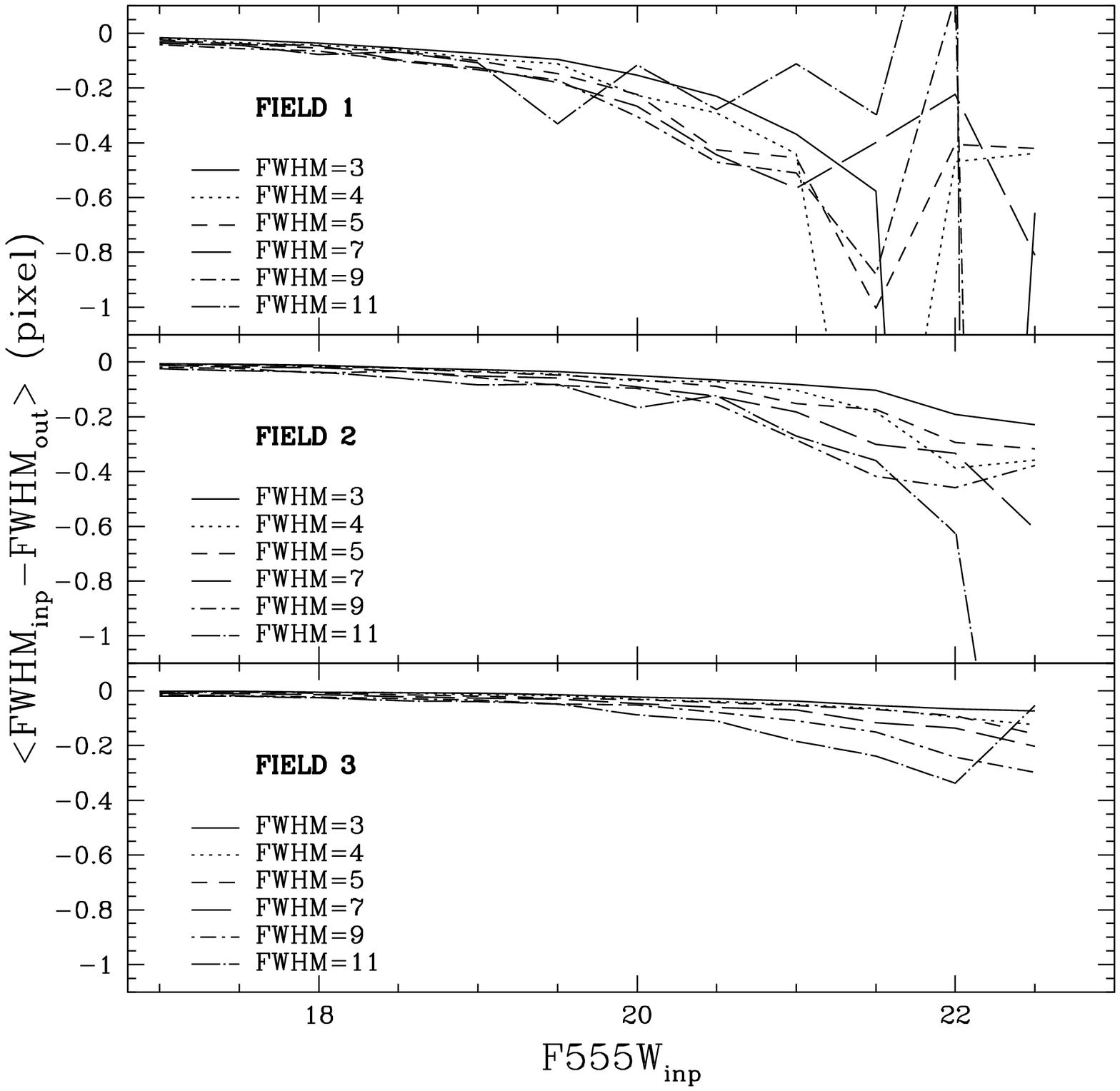}
\caption{Average input minus output FWHM (in pixels) for the artificial star clusters
as a function of the F555W input cluster magnitude,
for different FWHM input values and for the three test fields. \label{fwhmtest}
}
\end{figure}

\begin{figure}
\epsscale{1}
\plotone{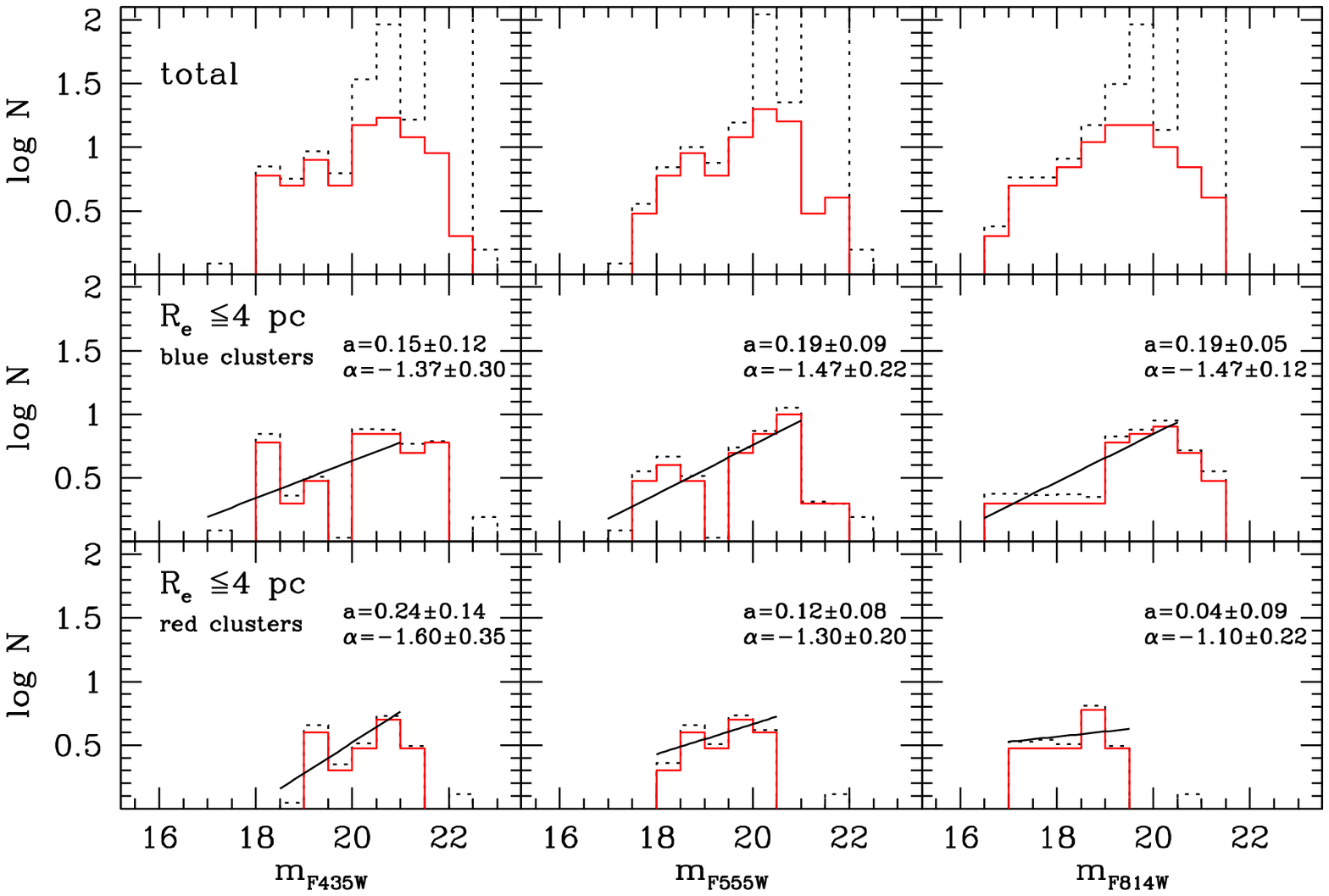}
\caption{Distributions of cluster magnitudes in the three bands (F435W, F555W, F814W).
The solid histogram is the observed distribution, while the dotted one is the completeness-corrected 
distribution. The histograms in the top row of panels refer to the total cluster sample, 
while in the middle and bottom panles we applied a cut in size of $R_e\leq$4 pc. 
The separate distributions for blue ($m_{F555W}-m_{F814W}<0.9$ or $m_{F435W}-m_{F555W}<0.5$) and red ($m_{F555W}-m_{F814W}\ge0.9$ and $m_{F435W}-m_{F555W}\ge0.5$) clusters are shown. The straight lines are the linear least squares fits 
in the form $\log N= a m + b$ to the different distributions, obtained considering only the magnitude bins brighter than 21.
At magnitudes fainter than this we are not able to properly correct the LFs (see Section~\ref{cluster_lf} for details).  
The slope a of the fit, and the slope $\alpha$ of the LF in the form  $dN(L)\propto L^{\alpha} dL$, are given in each panel. \label{lf}
}
\end{figure}

\begin{figure}
\epsscale{0.9}
\plotone{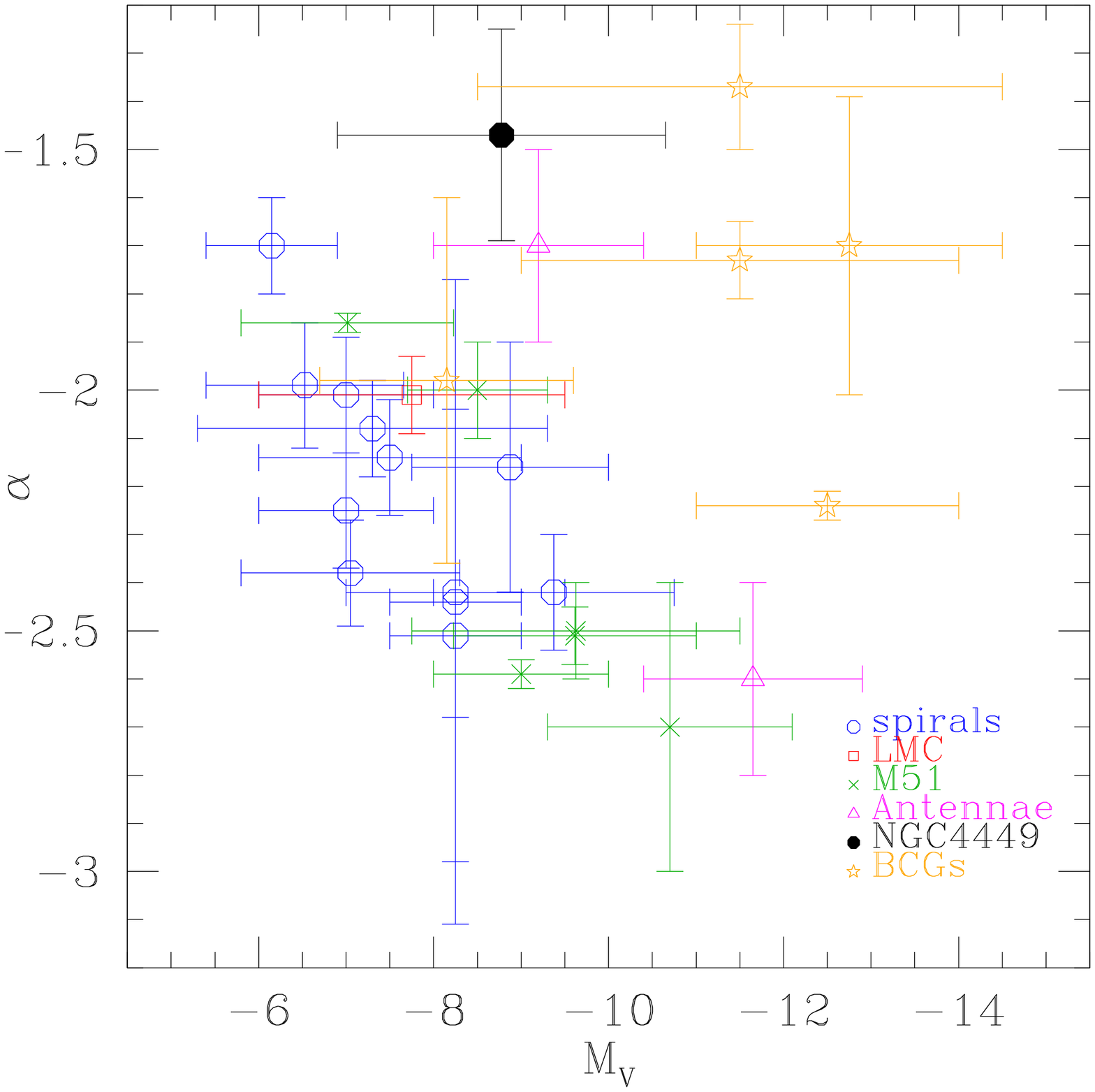}
\caption{Literature luminosity function slopes $\alpha$ in the V band as a function of the fit interval for young clusters in different galaxies:
spiral galaxies \citep[NGC~628, NGC~1313, NGC~3184, NGC~4395, NGC~5236, NGC~6744, NGC~6946, NGC~7793;][]{lar02,mora09}; 
the LMC \citep{lar02}; M~51 \citep{gieles06,haas08,hwang08}; the Antennae \citep{whitmore99};  
blue compact galaxies  \citep[Mrk930; ESO~185-IG13; NGC~1705,][]{adamo11,adamo11b,ann09}; NGC~4449 (this work). 
For NGC~1705, the leftmost BCG in the diagram, the slope is the average of the binned histogram fitting result  ($\alpha\sim-1.6$)  
and of the maximum likelihood estimator result  ($\alpha\sim-2.3$) obtained from the cluster magnitudes reported in \cite{ann09}. 
The significant difference between the two values is likely due to the small statistics. 
\label{alpha}
}
\end{figure}

\begin{center}
\begin{figure}
\epsscale{1.}
\plotone{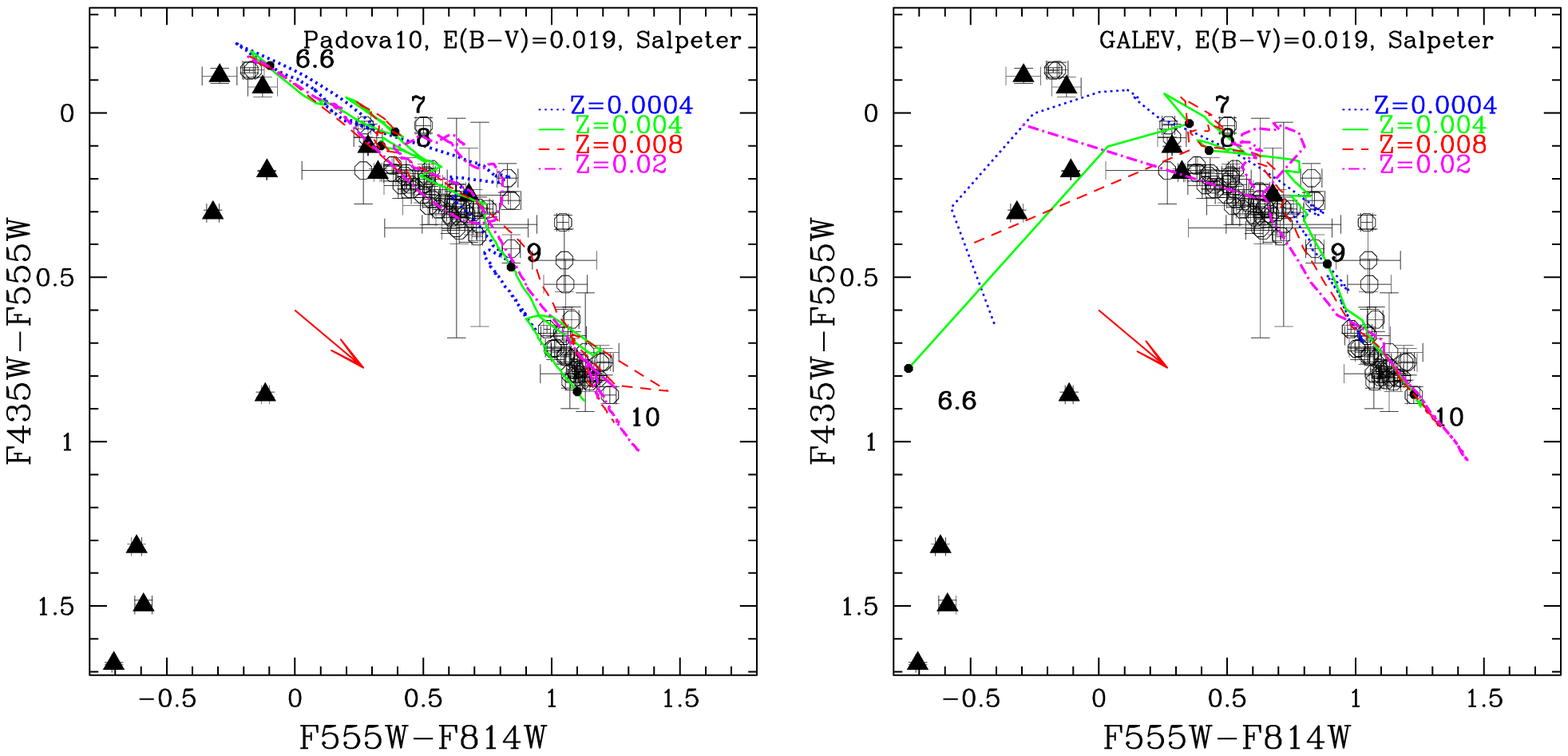}
\caption{F435W$-$F555W (B$-$V) versus F555W$-$F814W (V$-$I) color-color diagram 
for the candidate star clusters. Triangles indicate clusters with associated H$\alpha$ emission shown in Fig.~\ref{clu_ha}. 
Overplotted are the Padova simple stellar populations \citep{pad10} 
and the GALEV models (\cite{galev}, right panel) for a Salpeter IMF, a Galactic reddening 
E(B-V)=0.019, and different metallicities. For the Z$=0.004$ metallicity, we indicate the models of 
log(age[yr])$=$6.6, 7, 8, 9, and 10.  The arrow indicates the reddening vector for $E(B-V)=0.2$. \label{colcol}}
\end{figure}
\end{center}

\begin{center}
\begin{figure}
\epsscale{1.1}
\plotone{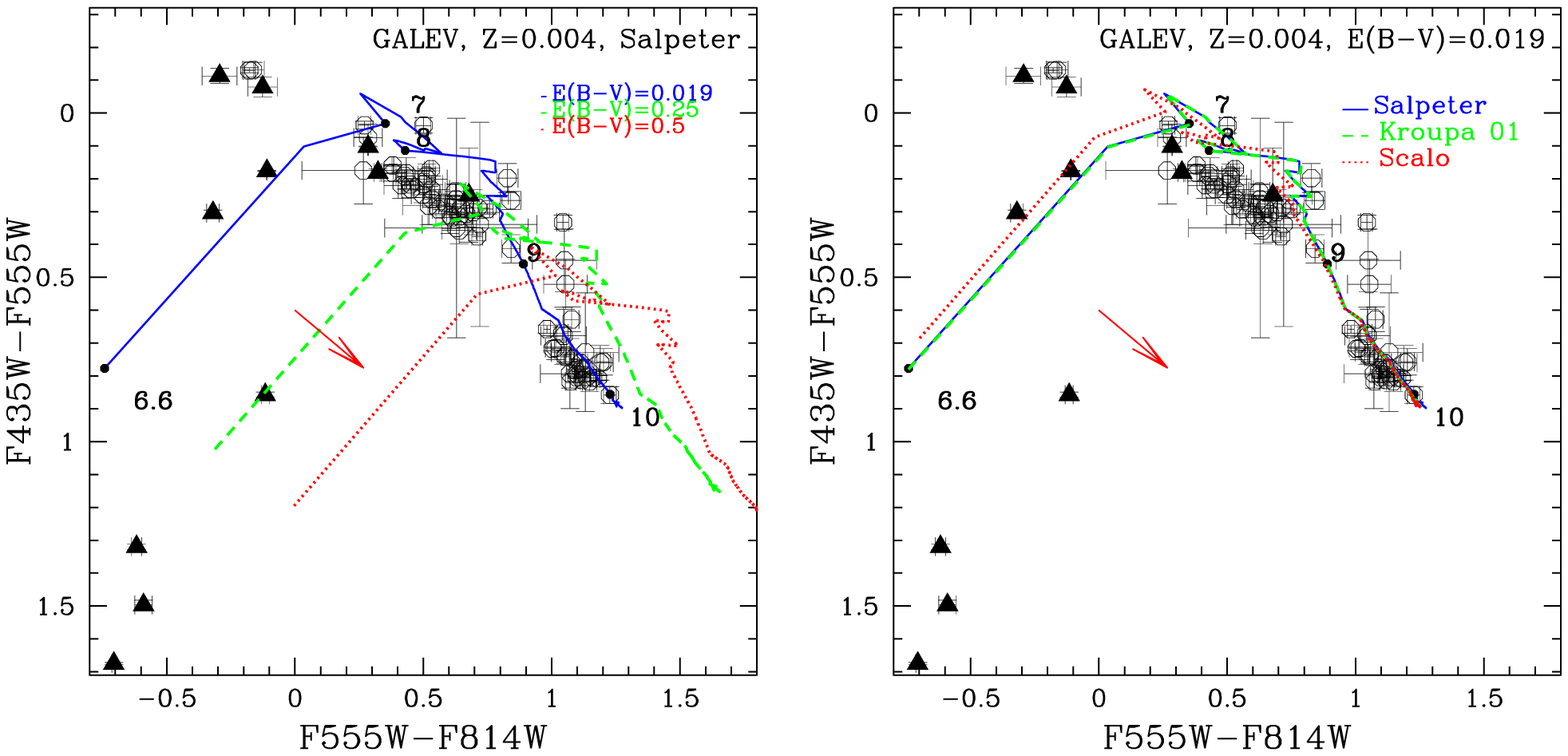}
\caption{F435W$-$F555W (B$-$V) versus F555W$-$F814W (V$-$I) color-color diagram 
for the candidate star clusters. Triangles indicate clusters with associated H$\alpha$ emission shown in Fig.~\ref{clu_ha}.  
Overplotted are the GALEV models (\cite{galev} 
for Z$=$0.004 and for different reddenings (left panel) and IMFs
 (right panel). Notice that the line corresponding to Kroupa' s IMF completely 
 overlaps Salpeter' s line in the right panel. For the Z$=0.004$ metallicity, we indicate the models of 
log(age[yr])$=$6.6, 7, 8, 9, and 10. The arrow indicates the reddening vector for $E(B-V)=0.2$. \label{colcol2}}
\end{figure}
\end{center}

\begin{center}
\begin{figure}
\epsscale{1.1}
\plotone{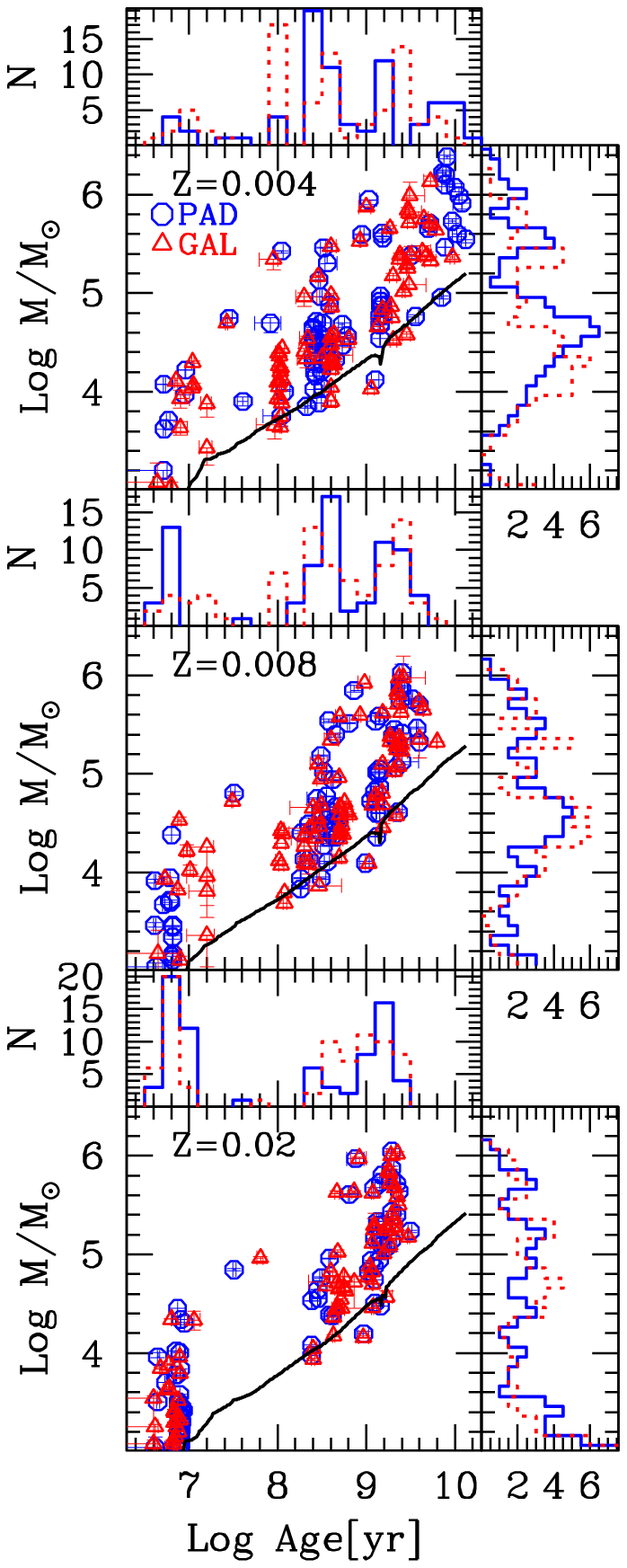}
\caption{Cluster masses versus ages derived for three 
metallicities (Z$=$0.004, 0.008, 0.02) with the Padova 
(\cite{pad10}, blue circles) and with the GALEV (\cite{galev}, red triangles) models.
The solid curve represents a cluster of magnitude $m_{F555W}=21.5$ 
at different ages at the distance of NGC~4449.  For each metallicity, 
the age and mass distributions are also shown (blue solid line for Padova, 
red dotted line for GALEV).
\label{agemass}}
\end{figure}
\end{center}

\begin{center}
\begin{figure}
\epsscale{1.1}
\plotone{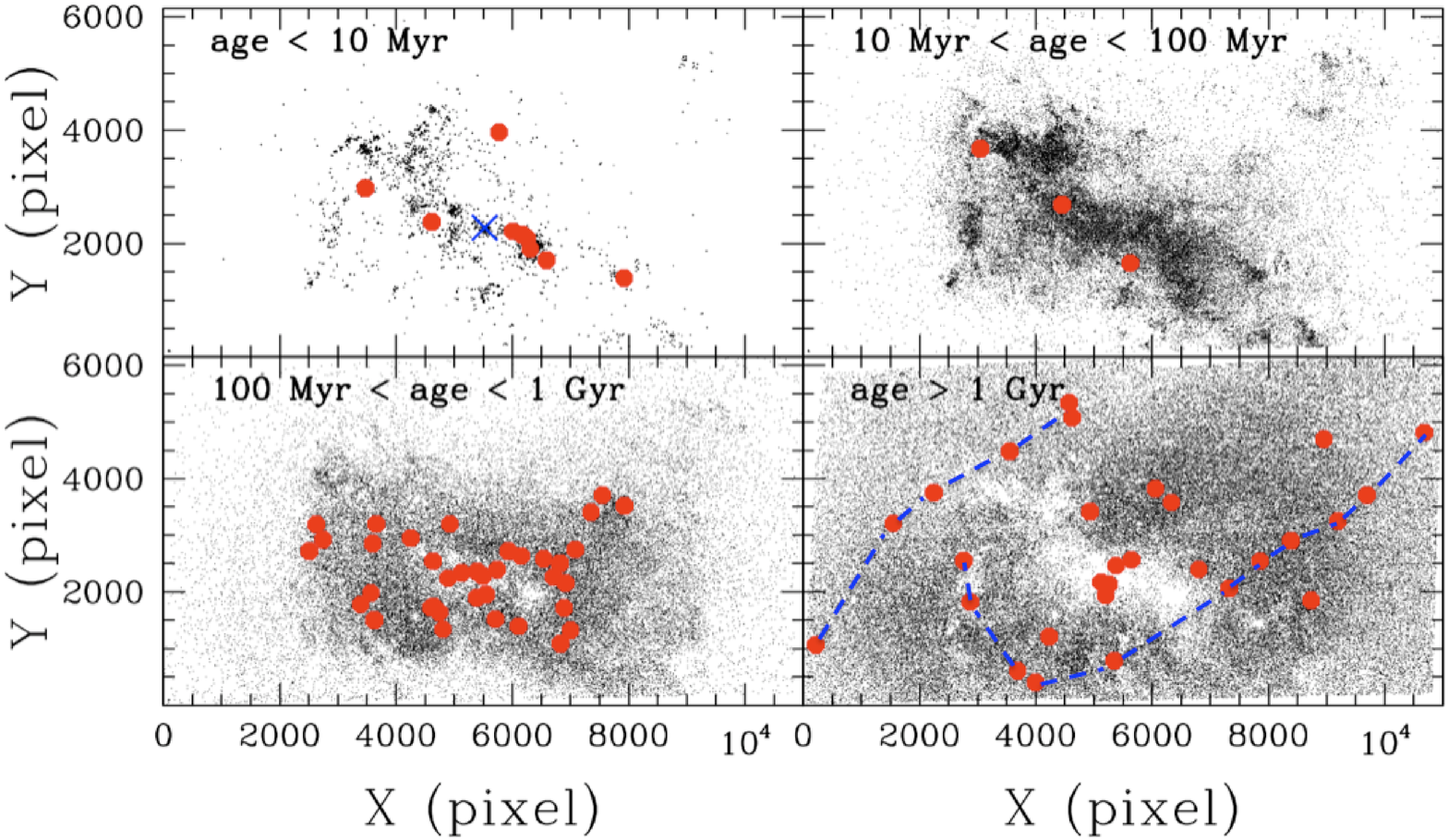}
\caption{Spatial distribution of stars (small dots) and clusters (big dots) 
in different age intervals. The cluster ages are those provided in Table~1, 
assuming  Z$=$0.004. For the stars, the selection  
was performed in \cite{ann08} through comparison of the ACS CMD 
with stellar evolutionary models at the same metallicity.
The cross indicates the position of the central super star cluster, which is 
saturated in our images, and for which we have assumed and age $\lesssim$10 Myr from \cite{bok01}. 
The dashed lines in the bottom right panel are drawn to guide the eye in identifying the ``linear structures''  
followed by the clusters.
\label{starclu}}
\end{figure}
\end{center}


\begin{center}
\begin{figure}
\epsscale{1.}
\plotone{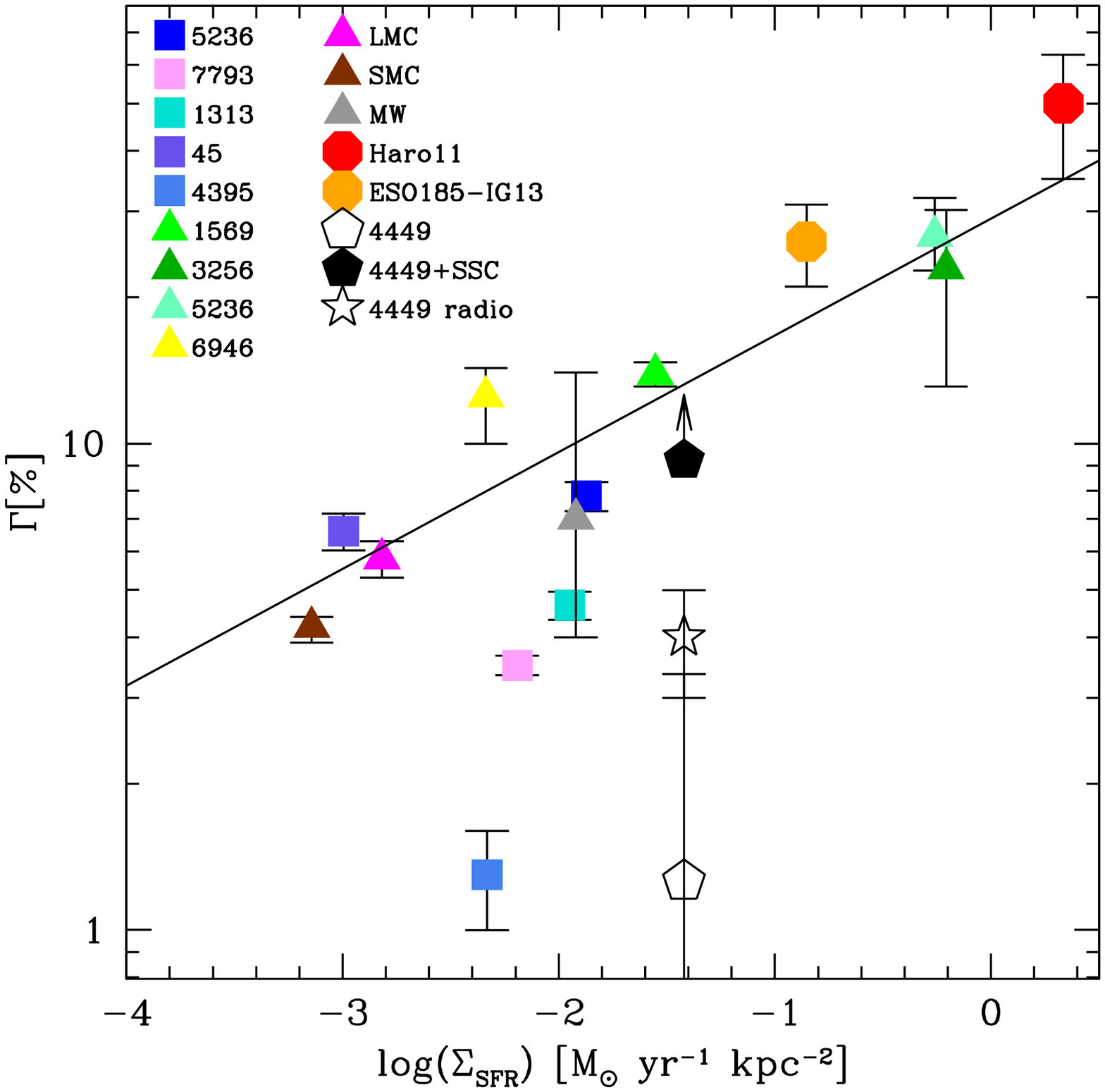}
\caption{Star formation fraction occurring in clusters ($\Gamma$) versus SFR density for galaxies from different samples: \cite{silva11}, squares 
(values obtained averaging Cols.~5  and 8 of Table~9); \cite{goddard10}, triangles; \cite{adamo10,adamo11b}, circles. Pentagons are the values for NGC~4449 from this work with (full) and without  (empty) the central SSC. The star symbol denotes the value obtained for NGC~4449 from the radio-detected embedded clusters from \cite{reines08}. The solid line is the relation by \cite{goddard10}.
 \label{gamma}}
\end{figure}
\end{center}

\begin{figure}
\epsscale{0.4}
\plotone{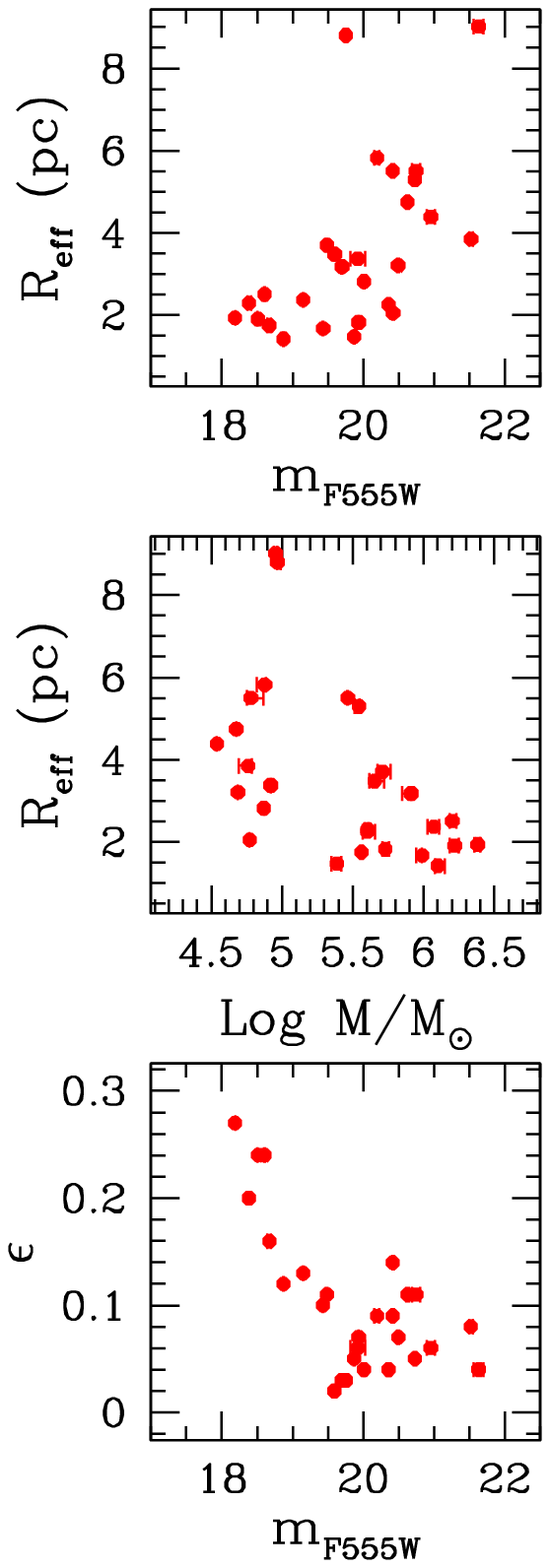}
\caption{For the red ($V-I>0.9$, $B-V>0.5$) clusters, effective radius $R_{e}$ versus F555W magnitude (top panel), 
 $R_{e}$ versus cluster mass (middle panel), and ellipticity $\epsilon$ versus 
 F555W magnitude (bottom panel). The plotted values are given in Table~1. \label{correla}}
\end{figure}

\begin{center}
\begin{figure}
\epsscale{0.95}
\plotone{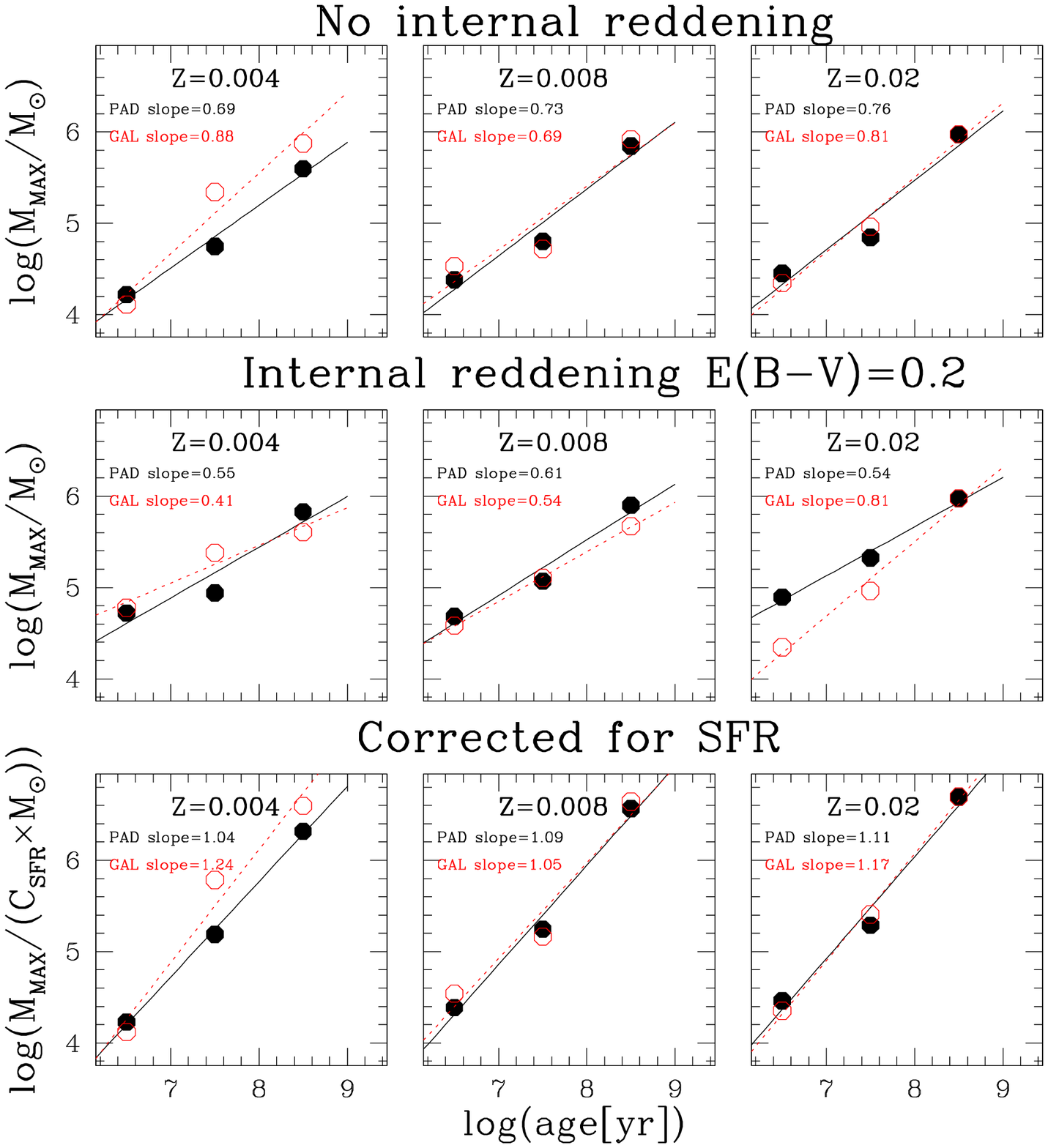}
\caption{Evolution of $\log(M_{max}$) with equal size (1 dex) $\log(age)$ bins for the
young (age$<$1 Gyr) clusters in NGC~4449, assuming different metallicities and adopting both the GALEV (\cite{galev}, open circles) 
and the Padova (\cite{pad10}, full circles) models. Fits to the $\log(M_{max}$) vs. $\log(age)$ 
for GALEV and Padova are shown as dotted and solid lines. 
Top panels: no internal extinction has been assumed. 
Central panels: we have assumed an internal extinction of $E(B-V)=0.2$. 
Bottom panels: the normalized  $\log(M_{max}/C_{SFR})$, where $C_{SFR}$ is proportional to the SFR 
in the different age bins, is plotted. No internal extinction is assumed. 
See Section~\ref{cluster_disruption} for details. \label{mmax_tot}}
\end{figure}
\end{center}

\end{document}